\documentclass[twocolumn,trackchanges]{aastex6}

\usepackage{amsmath}
\usepackage{array}
\newcolumntype{C}{>{\centering\arraybackslash}p{5em}}
\usepackage{colortbl}
\newcommand{\RowColor}{\rowcolor{gray!30}[\tabcolsep][\tabcolsep]\cellcolor{white}}
\newcommand{\mpc}{\cellcolor{white}}
\newcommand{\allc}{\cellcolor{blue!10}}

\newcommand{\nMWAlo}{90} 
\newcommand{\nMWAloMP}{79} 
\newcommand{\nMWAloIP}{11} 

\newcommand{\nMWAmi}{386} 
\newcommand{\nMWAmiMP}{336} 
\newcommand{\nMWAmiIP}{50} 

\newcommand{\nMWAhi}{648} 
\newcommand{\nMWAhiMP}{560} 
\newcommand{\nMWAhiIP}{88} 

\newcommand{\nMWAcont}{407} 
\newcommand{\nMWAcontMP}{341} 
\newcommand{\nMWAcontIP}{66} 

\newcommand{\nPKShi}{231} 
\newcommand{\nPKShiMP}{217} 
\newcommand{\nPKShiIP}{14} 

\newcommand{\nPKSlo}{6344} 
\newcommand{\nPKSloMP}{5306} 
\newcommand{\nPKSloIP}{1038} 

\newcommand{\alphamwalomimean}{-0.7} 	
\newcommand{\alphamwalomistd}{1.4}
\newcommand{\alphamwalomi}{-0.7\pm1.4}

\newcommand{\alphamipkslo}{-1.1\pm0.7}

\newcommand{\alphapkslohimean}{-2.6}	
\newcommand{\alphapkslohistd}{0.5}
\newcommand{\alphapkslohi}{-2.6\pm0.5}

\begin{document}
\title[]{Spectral Flattening at Low Frequencies in Crab Giant Pulses}


\author{B. W. Meyers\altaffilmark{1,2}, 
	S. E. Tremblay\altaffilmark{1}, 
	N. D. R. Bhat\altaffilmark{1}, 
	R. M. Shannon\altaffilmark{2}, 
	F. Kirsten,
	M. Sokolowski\altaffilmark{1},
	S. J. Tingay and
	S. I. Oronsaye\altaffilmark{1}}
\affil{International Centre for Radio Astronomy Research (ICRAR), Curtin University\\
	1 Turner Ave., Technology Park,\\ 
	Bentley, 6102, WA, Australia}
\and
\author{S. M. Ord}
\affil{Commonwealth Scientific and Industrial Research Organisation (CSIRO)\\
	Corner Vimiera \& Pembroke Roads\\
	Marsfield, 2122, NSW, Australia}

\email{bradley.meyers@postgrad.curtin.edu.au}

\altaffiltext{1}{ARC Centre of Excellence for All-Sky Astrophysics (CAASTRO)}
\altaffiltext{2}{CSIRO Astronomy and Space Science, PO Box 76, Epping, NSW 1710, Australia}

\begin{abstract}
We report on simultaneous wideband observations of Crab giant pulses with the Parkes radio telescope and the Murchison Widefield Array (MWA).
The observations were conducted simultaneously at 732 and 3100\,MHz with Parkes, and at 120.96, 165.76 and 210.56\,MHz with the MWA.
Flux density calibration of the MWA data was accomplished using a novel technique based on tied-array beam simulations.
We detected between \nMWAlo--\nMWAhi\ giant pulses in the 120.96--210.56\,MHz MWA subbands above a $ 5.5\sigma $ threshold while in the Parkes subbands we detected \nPKSlo\ and \nPKShi\ giant pulses above a threshold of $ 6\sigma $ at 732 and 3100\,MHz, respectively.
We show, for the first time over a wide frequency range, that the average spectrum of Crab giant pulses exhibits a significant flattening at low frequencies.
The spectral index, $ \alpha $, for giant pulses evolves from a steep, narrow distribution with a mean $ \alpha=\alphapkslohimean $ and width $\sigma_\alpha=\alphapkslohistd $ between 732 and 3100\,MHz, to a wide, flat distribution of spectral indices with a mean $ \alpha=\alphamwalomimean $ and width $ \sigma_\alpha=\alphamwalomistd $ between 120.96 and 165.76\,MHz. 
We also comment on the plausibility of giant pulse models for Fast Radio Bursts based on this spectral information.
\end{abstract}


\section{INTRODUCTION} \label{sec:introduction}
The Crab pulsar (PSR J0534+2200) was discovered through its giant pulse emission \citep{1968SciStaelin}.
Giant pulses are short-duration bursts of emission, lasting for $ \lesssim 1\mathrm{\,ns} $ to $ \sim 10\mathrm{\,\mu s} $, that appear only within a small fraction of the normal pulse phase window \citep{2003NatHankins,2007AAPopov,2008ApJBhat}.
Individual giant pulses are observed to have brightness temperatures in the range $ T_\mathrm{b}\sim10^{30\text{--}32}\mathrm{\,K} $, implying a coherent emission mechanism.
At extremely high time resolution, Crab giant pulses have been observed to reach brightness temperatures of $ 10^{41}\mathrm{\,K} $, corresponding to a peak flux density of $ S_\mathrm{peak}=2.2\mathrm{\,MJy} $ at $ 9\mathrm{\,GHz} $ \citep{2007ApJHankins}.
Giant pulses are therefore invaluable tools for understanding pulsar emission and, more generally, astrophysical coherent emission mechanisms from a variety of objects.

It has been established that the occurrence of giant pulse energies follow a power-law distribution (e.g. \citealp{1972ApJArgyle,2004ApJCordes,2008ApJBhat,2015ApJOronsaye}), while normal pulse energies tend to exhibit an exponential or log-normal distribution (e.g. \citealp{2012MNRASBurke-Spolaor}).
There are six pulsars known to exhibit giant pulses.
These include: two young pulsars (PSRs J0534+2200 and J0540-6919) and four millisecond pulsars (PSRs J0218+4232, J1823-3021A, J1824-2452A, J1939+2134; \citealp{2006ApJKnight}), all of which have high magnetic field strengths at the light-cylinder radius ($ B_\mathrm{LC} \sim 10^{5\text{--}6}\mathrm{\,G} $).
The giant pulses from these six objects occur within a confined phase location, are intrinsically short duration (microseconds or less) and exhibit a power-law pulse energy distribution.
In the literature, there are several other pulsars which emit large amplitude pulses, often referred to as ``giant pulses'' (e.g. B0950+08; \citealp{2012AJSingal,2015AJTsai,2016AJTsai}, and J1752+2359; \citealp{2005AAErshov}). 
It is not necessarily clear if the emission from these pulsars shares the distinctive characteristics exhibited by the above six confirmed cases.

The physics responsible for producing these coherent bursts of radio emission is unknown, but is thought to be a broadband, non-linear plasma process (e.g. \citealp{2016JPlPhEilek,2016JPlPhMelrose}) that is able to produce detectable emission from radio to $ \gamma $-ray frequencies (e.g. \citealp{2010ApJAbdo}).
While Crab giant pulses appear to be a broadband phenomenon, detectable across the full observing bandwidth in most observations, they are not expected to always be detected simultaneously over multiple widely separated frequency bands (e.g. \citealp{1999ApJSallmen,2015ApJOronsaye}).

The flux density spectrum of normal pulsar emission is typically described by a simple power-law model $ S_\nu\propto \nu^\alpha $ where $\langle\alpha\rangle = -1.8\pm0.2$, for observing frequencies $>100$\,MHz (e.g. \citealp{1973AASSieber,1995MNRASLorimer,2000AASMaron}).
The underlying distribution of pulsar spectral indices, based on Monte Carlo simulations of pulsar surveys, has also been modeled as a Gaussian distribution with a mean of $ \langle\alpha\rangle = -1.4\pm1 $ \citep{2013MNRASBates}.
Only a handful of cases ($ \sim 10\% $) are known where a different spectral shape is observed, such as a broken power-law or flat spectrum.
There are also the peculiar ``gigahertz peaked spectra'' (GPS) pulsars \citep{2007AAKijak,2011AAKijak}, where the spectrum peaks and turns over at $\sim 1$\,GHz.
The spectral shape of these GPS pulsars is believed to be a consequence of the pulsar local environment (e.g. \citealp{2012ASPCDembska,2016MNRASRajwade}).
For giant pulses, spectral flattening or a turn-over has not yet been directly observed.
\citet{2015ApJOronsaye} suggested, via Monte-Carlo analysis, that there was a $ \sim 5\% $ flattening of the spectral index distribution mean between 193\,MHz and 1382\,MHz.
More simultaneous, wideband observations are therefore necessary to constrain the spectral behavior of giant pulses, both individually and statistically for the population.

Multi-frequency simultaneous observations of the Crab have previously been undertaken, though typically only between two frequencies (e.g. \citealp{2008ApJBhat,2015ApJOronsaye}) or over a narrow frequency range (e.g \citealp{2012AAKaruppusamy,2016ApJEftekhari}).
In order to further constrain the giant pulse emission mechanism, wideband simultaneous observations with intermediate frequency coverage such as that conducted by \citet{2016ApJMikami} are required to uncover the broadband spectral behavior.

With the advent of the Fast Radio Burst (FRB) phenomenon, especially the repeating FRB 121102 \citep{2014ApJSpitler,2016NatSpitler,2016ApJScholz}, several theories have been put forth suggesting that at least some FRBs may originate from extragalactic giant pulses (e.g. \citealp{2016MNRASCordes,2016MNRASLyutikov,2016MNRASConnor}).
Determining the spectral behavior of simultaneously detected Crab giant pulses over a wide frequency range will also provide clues regarding a giant pulse origin of FRBs, especially given the paucity of low frequency detections. 

In this article, we report on simultaneous observations of giant pulses from the Crab pulsar conducted with the Parkes radio telescope and the Murchison Widefield Array (MWA; \citealp{2013PASATingay}).
The Parkes 64-m radio telescope is well-known for pulsar science and facilitated our high-frequency observations (732\,MHz and 3.1\,GHz).
The MWA is a low-frequency (70--300\,MHz) Square Kilometre Array precursor located in Western Australia at the Murchison Radio-astronomy Observatory.
With the high time resolution Voltage Capture System (VCS; \citealp{2015PASATremblay}), the MWA provided our low-frequency observations. 
We present the detection and analysis of simultaneous giant pulses between the MWA and the Parkes radio telescope, covering 120--3100\,MHz with 1--3 intermediate observing bands.

This paper is organized as follows.
In Section~\ref{sec:observations} we describe the setup for the MWA and Parkes observations, and in Section~\ref{sec:processing_calibration} we describe the post-processing and data calibration.
Section~\ref{sec:results} describes the methods used to detect simultaneous giant pulses from both instruments, and details the results of the analysis focusing on giant pulse spectra.
In Section~\ref{sec:discussion} we discuss the implications of our results for the giant pulse emission mechanism and briefly comment on the applicability of a giant pulse model to explain the emission observed from FRBs.
We summarize and conclude in Section~\ref{sec:conclusion}.
Throughout, we adopt $ S_\nu $ to represent flux densities and $ F_\nu $ to represent fluences at frequency $ \nu $.

\section{OBSERVATIONS}\label{sec:observations}
The Crab pulsar was observed simultaneously with Parkes and the MWA on 7 November 2014.
Parkes observed the pulsar at 732 and 3100\,MHz for 1.4 hours.
The MWA-VCS data collection was split into two distinct observations, totaling 1.3 hours.
The first 20 minute observation was conducted at a central frequency of 184.96\,MHz.
Immediately following this, the second observation lasted for 1 hour and was designed such that the MWA bandwidth was split into four subbands distributed between 120.96--278.40\,MHz.
Observation details are summarized in Table~\ref{tab:observations}.
\begin{table*}
\centering
\caption{Observation parameters.\label{tab:observations}}
\begin{tabular}{l|cccc|cc}
\hline Parameters & \multicolumn{4}{c|}{MWA\tablenotemark{$a$}} & \multicolumn{2}{c}{Parkes} \\ 
\hline\hline
Center frequency (MHz) & 120.96 & 165.76 & 184.96 & 210.56 & 732 & 3100 \\ 
Bandwidth (MHz) & 7.68 & 7.68 & 30.72 & 7.68 & 64 & 1024 \\
FWHM (arcmin) & 3.60 & 2.63 & 2.36 & 2.07 & 26.39 & 6.23 \\ 
Time resolution ($ \mathrm{\mu s} $) & 100 & 100 & 100 & 100 & 256 & 256 \\ 
Frequency resolution (MHz) & 0.01 & 0.01 & 0.01 & 0.01 & 0.125 & 2 \\
Dispersion delay across bandwidth\tablenotemark{$b$} (ms) & 2048.48 & 795.26 & 2319.09 & 387.83 & 77.16 & 17.11 \\
Dispersion delay in lowest channel\tablenotemark{$b$} (ms) & 2.93 & 1.10 & 0.96 & 0.53 & 0.17 & 0.05 \\
Start time (UTC) & 17:14:00 & 17:14:00 & 16:53:40 & 17:14:00 & 16:48:27 & 16:48:37 \\
Observation duration (s) & 3663 & 3663 & 1163 & 3663 & 5065 & 5055 \\
\hline 
\end{tabular} 
\tablenotetext{a}{The 278.4\,MHz subband was excluded due to poor quality calibration solutions (see text).}
\tablenotetext{b}{Assuming a nominal dispersion measure of $56.7762\mathrm{\,pc\,cm^{-3}}$.}
\end{table*}

\subsection{Parkes}\label{sec:parkes_obs}
We observed the Crab pulsar using the coaxial 1050cm receiver on the 64-m Parkes radio telescope, which is capable of simultaneously recording signals at 732\,MHz (64\,MHz bandwidth) and 3100\,MHz (1024\,MHz bandwidth). 
Both systems are sensitive to linear polarization.
Data were recorded with the mark-3 and mark-4 versions of the Parkes digital filterbank spectrometers (PDFB3 and PDFB4) for a duration of $ \approx 5060\mathrm{\,s} $. 
The spectrometers employ polyphase digital filters, with PDFB3 recording data with 512 channels across the 64\,MHz low-frequency band, and PDFB4 recording data with 512 channels across the 1024\,MHz high-frequency band.
Data were recorded in polarimetric search mode.
For each channel four coherency products (the power from each probe and the complex-valued correlated power between the two) were detected and averaged over 256\,$\mu$s before being written to disk with 8-bit precision.

The decorrelation bandwidths ($ \Delta\nu_\mathrm{DISS} $) due to diffractive scintillation at 732 and 3100\,MHz are 35\,kHz and 6\,MHz respectively, assuming $ \Delta\nu_\mathrm{DISS}=2.3\mathrm{\,MHz} $ at $ 2.33\mathrm{\,GHz} $ \citep{2004ApJCordes} and a scaling of $ \Delta\nu_\mathrm{DISS}\propto\nu^{3.6} $ (e.g. \citealp{2013ApJEllingson,2016ApJEftekhari}, Kirsten et al. in prep.).
Over the respective bandwidths of the observing frequency bands, these contributions are negligible.
The refractive time scales are 2 days and $ \sim 7 $ hours respectively.
On the time scales we are probing, we do not expect any significant contribution from scintillation to the giant pulse flux densities in the 732\,MHz band.
In the 3100\,MHz band we expect that the small contribution from scintillation will be dominated by the measurement scatter in giant pulsar flux densities.

\subsection{Murchison Widefield Array}\label{sec:mwa_obs}
The MWA is a low-frequency array composed of 128 tiles, with each tile consisting of 16 dipoles evenly spaced in a regular $4\times 4$ meter grid.
The MWA has 30.72\,MHz instantaneous bandwidth that can be separated into 24 independent 1.28\,MHz subbands, which can be distributed across the $ 70\text{--}300\mathrm{\,MHz} $ observing range. 

The VCS is the high-time and frequency resolution observing system for the MWA, capable of capturing the tile voltages after the channelization stage within the MWA signal processing pipeline.
This allows critically sampled complex tile voltages (100\,$\mu$s time resolution, 10\,kHz frequency resolution) to be recorded to on-site disks at a data rate of $\sim 28$\,TB per hour.
Using the VCS, we recorded $ \approx 4826\mathrm{\,s} $ of data from the array pointed towards the Crab pulsar (see Table~\ref{tab:MWA_efficiencies}).
As previously mentioned, this observing run was split into two observations.
The first 20 minutes with the full 30.72\,MHz of bandwidth, centered at 184.96\,MHz. 
The remaining 60 minutes were observed with the bandwidth split into four 7.68\,MHz subbands, distributed to center frequencies of 120.96, 165.76, 210.56 and 278.40\,MHz.

At MWA frequencies, the decorrelation bandwidths due to diffractive scintillation range between $ \Delta\nu_\mathrm{DISS}\approx 50\text{--}1000\mathrm{\,Hz} $ at the observed MWA bands.
The refractive time scales are between 8 and 25 days.
Therefore, we do not expect any contribution from scintillation to be significant in our intensity estimates for the Crab giant pulses at MWA frequencies.

\section{DATA PROCESSING AND CALIBRATION}\label{sec:processing_calibration}

\subsection{Parkes}\label{sec:pks_calibration}
Absolute flux density calibration was performed by observing the radio galaxy Hydra A (3C 218) as part of the Parkes Pulsar Timing Array (PPTA) project \citep{2013PASAManchester}.
Polarization calibration was conducted by injecting a linearly polarized signal into the feeds. 
This allowed us to measure the frequency-dependent differential gain and phase of the two feeds.   
We did not correct for feed ellipticity or cross coupling.
  
The 732\,MHz data were incoherently dedispersed and folded using \textsc{dspsr} \citep{2011PASAvanStraten} with an ephemeris from the Jodrell Bank monthly monitoring\footnote{http://www.jb.man.ac.uk/pulsar/crab.html}.
A more accurate pulsar ephemeris was produced from these data, fitting for the optimum period, period derivative and dispersion measure with \textsc{tempo2} \citep{2006MNRASHobbs}.
The dispersion measure calculated by this process was $ 56.7762\mathrm{\,pc\,cm^{-3}} $ and is henceforth taken as the nominal dispersion measure for the Crab pulsar.  

Data from both Parkes bands were then re-processed using \textsc{dspsr} and the updated ephemeris, subdividing the data streams into individual pulses. 
The pulses were flux density and polarization calibrated using \textsc{psrchive} \citep{2004PASAHotan} routines.
RFI was removed using the \textsc{paz} routine, flagging the edge 5\% of each band and running the inbuilt median smoothed difference excision algorithm.

\subsection{Murchison Widefield Array}\label{sec:mwa_calibration}
Calibrating the MWA data is non-trivial, especially in the case of VCS recorded data for which there is currently no dedicated automatic calibration pipeline.
The Crab nebula was selected as the calibrator source for both the 184.96\,MHz full-bandwidth observation and the split-bandwidth observation.
Visibilities for each observation were created using an offline version of the MWA correlator (which performs the same function as the online version; \citealp{2015PASAOrd}).
For each band, a calibration solution (amplitude and phase) for each tile was calculated from the visibilities using the Real Time System (RTS; \citealp{2008ISTSPMitchell}).
The output from the RTS is a calibration solution for each coarse channel containing the calibration information for each MWA tile, thus there is a set of 24 solutions per observation.
Due to poor quality calibration solutions, data from 8 of the 128 tiles for the full bandwidth observation were discarded, while 21 of the 128 tiles were discarded for the split-bandwidth observation.

\begin{figure}
\includegraphics[width=\linewidth]{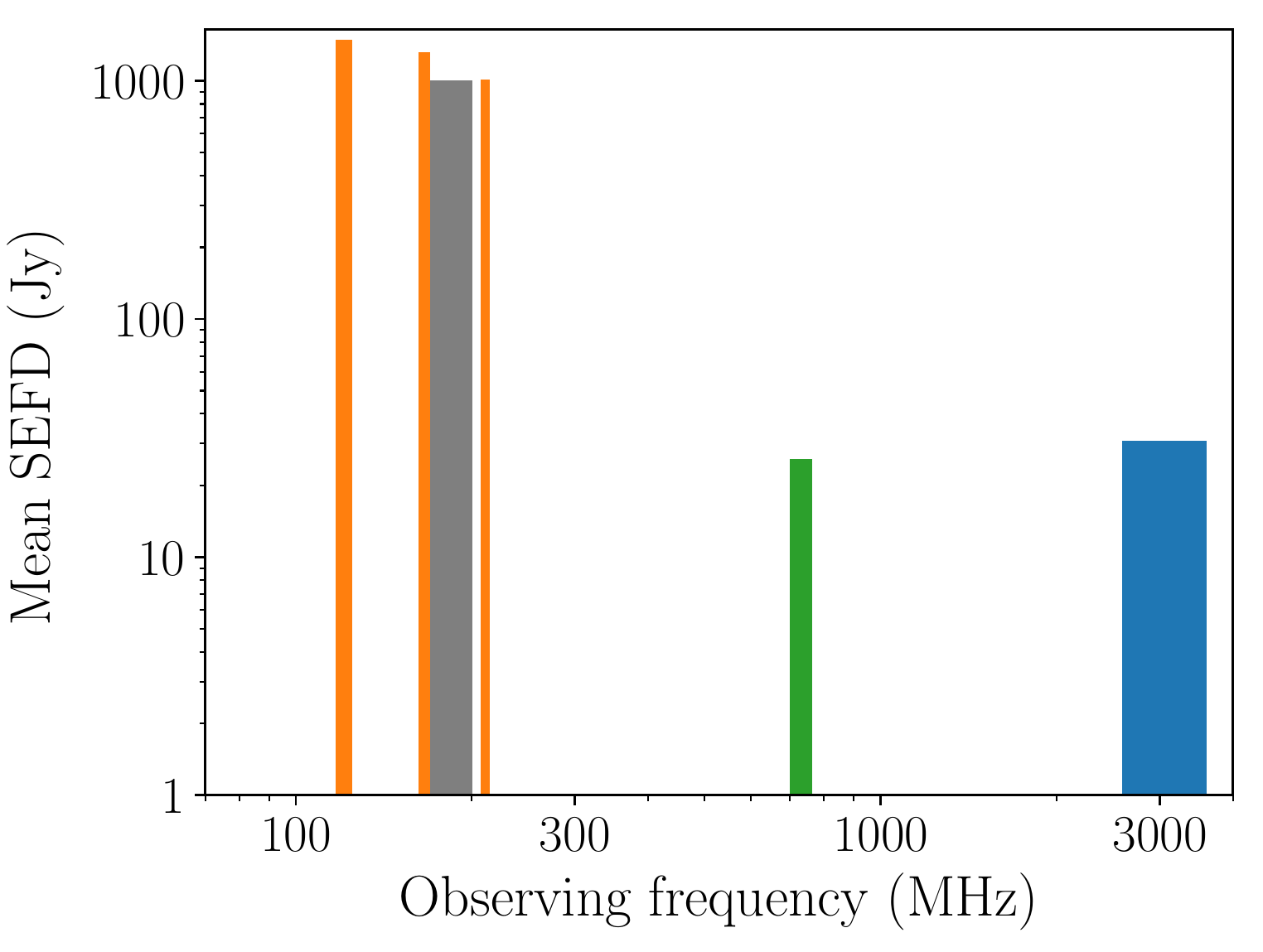}	
\caption{Schematic of the MWA and Parkes frequency coverage versus the mean system equivalent flux density (SEFD).
The orange bars correspond to the split-bandwidth observations with 7.69\,MHz bandwidth.
The gray is the full-bandwidth observation with 30.72\,MHz bandwidth.
The green bar represents the 732\,MHz Parkes band with 64\,MHz bandwidth and the blue bar represents the 3100\,MHz Parkes band with 1024\,MHz bandwidth.}
\label{fig:mwa_pks_bands}
\end{figure}

The MWA tiles and beam models are less well characterized at higher frequencies ($ \nu \sim 300 $\,MHz) and moreover there are increased levels of satellite-based radio frequency interference (RFI), making calibration significantly more difficult.
Owing to the poor calibration solution quality at the 278.40\,MHz band, the data were discarded leaving us with three usable subbands (120.96--210.56\,MHz) and one band at 184.96\,MHz (see Figure~\ref{fig:mwa_pks_bands}).

The MWA uses analogue beamformers to set the pointing direction of each tile, thus there are a discrete set of delays available.
For our observations, this means that the tile beam is never pointed directly at the Crab and so we are never at full sensitivity. 
The MWA tile beam is very complex, thus in some cases the Crab is not within a well-understood region of the beam.
Throughout the 120.96 and 165.76\,MHz and 184.96\,MHz observations, the Crab is always within the half-power point of the beam, for which we have the most confidence in the beam modelling. 
At 210\,MHz, the beam is such that the Crab is only barley within the half-power point for $ \sim 1/3 $ of the full observation.  
We therefore have less confidence in the ability to accurately flux calibrate the data at that particular frequency band, using the method outlined here.

\subsubsection{Tied-array beamforming}\label{sec:mwa_coherent_bforming}
The tied-array beam is formed by coherently summing individual tile voltages (i.e. sum tiles in phase and then detect power).
Theoretically, this process yields a factor of $ \sqrt{N_\mathrm{co}} $ improvement in sensitivity over an incoherent sum (i.e. detect tile power and sum, see \citealp{2015ApJOronsaye}), where $ N_\mathrm{co} $ is the number of tiles used to create the tied-array beam.
This corresponds to a potential order of magnitude increase in sensitivity for the MWA.
In reality, this is not the case and we see an improvement by a factor of between $ 4.2\text{--}5.4 $, depending on the frequency.
The discrepancy is primarily due to the pointing of the telescope (i.e. the MWA beam pattern is less well characterized as we diverge from a zenith pointing) and the calibration solution quality.

For MWA-VCS data, a tied-array (coherent) beam is created by a post-process beamforming pipeline (Ord et al. in prep.) implemented on the Galaxy cluster at the Pawsey Supercomputing Centre\footnote{https://www.pawsey.org.au/}.
The coherent beamforming pipeline involves incorporating the individual tile polarimetric response, both cable and geometric delay models, and complex gain information (amplitude and phase) for each tile, per frequency channel, based on the calibration solutions.
The tile weights used to create the tied-array beam are effectively determined by solving for the minimum $ \chi^2 $-error between the target data and the calibration model from the solutions.

\subsubsection{Tied-array system temperature and gain}\label{sec:mwa_gain_tsys}
For a tied-array beam, the field-of-view is significantly smaller than that of the tile beam, approximating the naturally weighted synthesized beam of the array -- nominally $ \mathrm{FWHM}\sim 1.27\lambda/D $, where $ \lambda $ is the observing wavelength and $ D $ is the maximum baseline of the array.
The scaling factor of 1.27 derives from the MWA being dominated by shorter baselines. 
This means that neither the integrated sky temperature nor the system gain will be the same as for the tile beam.
  
The overall system temperature ($ T_\mathrm{sys} $), for each frequency band, is a combination of the receiver temperatures ($ T_\mathrm{rec} $), antenna temperatures ($ T_\mathrm{ant} $), and the ambient temperature ($ T_0 $), and is calculated as
\begin{equation}
T_\mathrm{sys}=\eta T_\mathrm{ant}+(1-\eta)T_0 + T_\mathrm{rec},
\label{eq:tsys}
\end{equation}
where $ \eta $ is the frequency and direction dependent radiation efficiency of the array. 
Efficiencies and receiver temperatures for each subband are given in Table~\ref{tab:MWA_efficiencies}.
The receiver temperatures are well characterized across the nominal observing frequency range of the MWA. 
The ambient temperature weighting of $ 1-\eta $ where $ \eta \simeq 1 $ means that the contribution is negligible compared to the sky, and we therefore assume the ambient temperature is $ T_0\approx 290\mathrm{\,K} $.
This contributes $ \simeq 5\text{--}7\mathrm{\,K} $ to the total system temperature.

\begin{table}[!htbp]
\centering
\caption{Frequency and direction dependent radiation efficiencies and receiver temperatures for the MWA.\label{tab:MWA_efficiencies}}
\begin{tabular}{cccc}
\hline  
Pointing center & Center    & Radiation  & Receiver \\ 
(Az., El.)      & frequency & efficiency & temperature \\
(deg, deg)      & (MHz)     & $ \eta $   & $ T_\mathrm{rec} $ (K)\\
\hline\hline
(18.43, 41.42) & 120.96 & 0.980 & 39\\
(18.43, 41.42) & 165.76 & 0.976 & 32\\
(26.56, 37.31) & 184.96 & 0.980 & 23\\
(18.43, 41.42) & 210.56 & 0.981 & 34\\
\hline 
\end{tabular}
\end{table}

In order to calculate the antenna temperature in equation~(\ref{eq:tsys}) we require an adequate understanding of the tied-array synthesized beam pattern. 
In this case, the tied-array beam power pattern is the product of an individual MWA tile power pattern and the array factor. 
The tile pattern is simulated using the formalism set out by \citet{2015RaScSutinjo}, while the array factor encapsulates the phase information required to point the tied-array at the target source.
For a full description of the formulation of the array factor, see Appendix~\ref{appendix:arrayfactor}.
This procedure was used to create the tied-array beam pattern at multiple times throughout the observation.

We use the Global Sky Model (GSM; \citealp{2008MNRAS_GSM}) as our sky map and scale it to our observing frequencies.
The GSM was modified in the region of the Crab nebula with the scaling $ S_\mathrm{CN} = 955\nu^{-0.27}\mathrm{\,Jy} $ \citep{1973Ap&SSApparao,1997ApJBietenholz} to more accurately represent the contribution from the nebula.
Convolving the tied-array beam pattern with the GSM and integrating over the sky (see e.g. \citealp{2015PASASokolowski}), we produce an estimate of the antenna temperature (see Appendix~\ref{appendix:1:tsys_calc}).
Using these antenna temperatures and equation~(\ref{eq:tsys}), we calculate a system temperature estimate multiple times during the observation for each band.
Fitting a second-order polynomial to the results from the separate evaluations of $ T_\mathrm{sys} $, we estimate a system temperature curve as a function of time.

\begin{figure}
\includegraphics[width=0.48\textwidth]{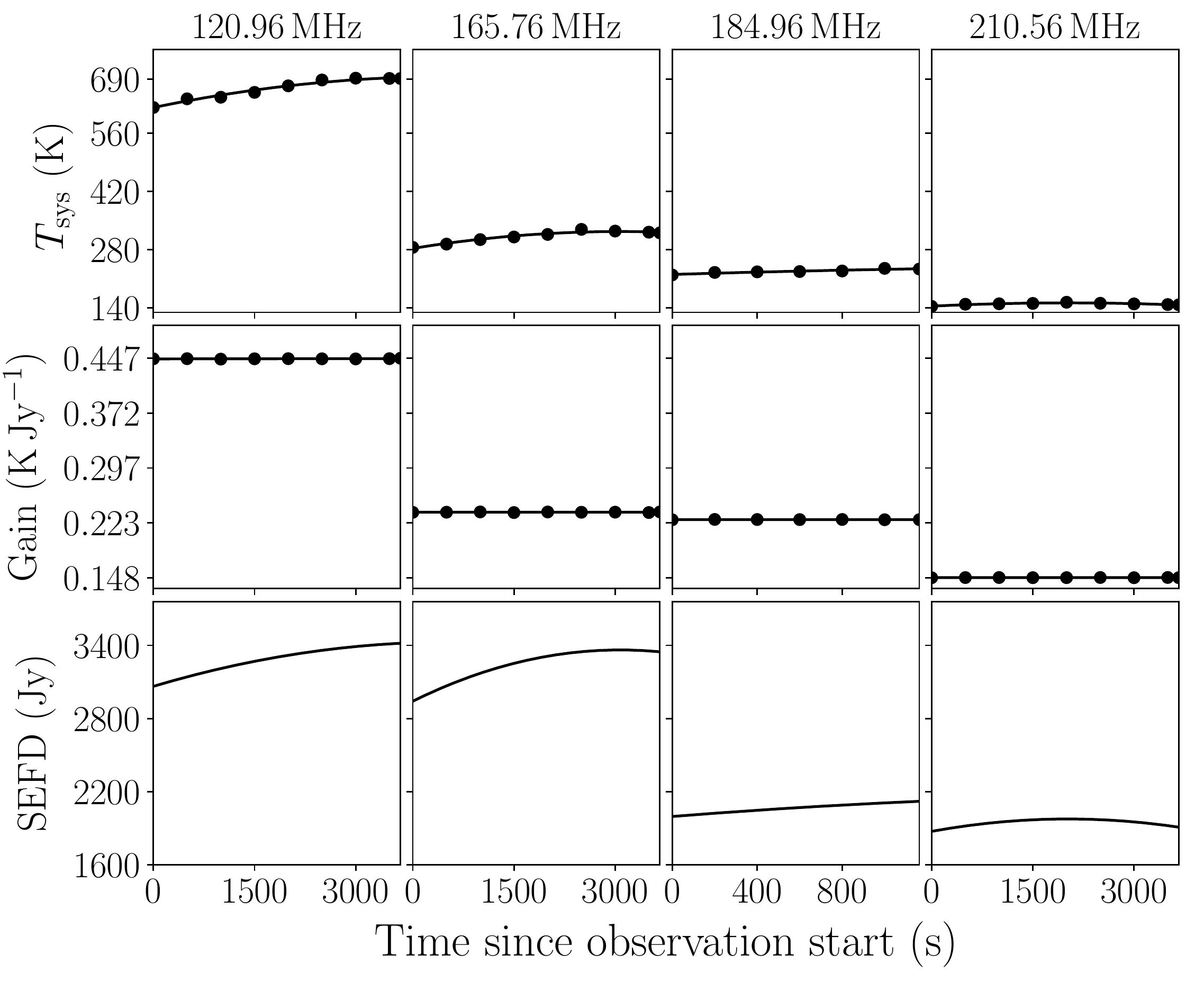}	
\caption{System temperature, gain and SEFD estimates as a function of time for the full-bandwidth and split-bandwidth MWA observations.
The black points are the measurements made from the simulated tied-array beam patterns and the black lines are fits to the measurements.
The system temperatures are fitted with a second order polynomial and the gains described by a linear fit.
The SEFD is calculated as $ f_cT_\mathrm{sys}/G $, using the polynomial fits, and $ f_c $ is defined in equation~\eqref{eq:coherency}.}
\label{fig:mwa_sefd}
\end{figure}

We also calculate the gain, $ G $, (see Appendix~\ref{appendix:2:gain_calc}) at the same intervals as calculating the system temperature.
The gains are relatively stable over the duration of the observation, thus we fit a linear slope to create a gain curve as a function of time for the entire observation. 
The system temperature and tied-array gain curves are shown in Figure~\ref{fig:mwa_sefd}.
Note that these estimates have included in them the assumption of ideal sensitivity increase (i.e. by a factor of $ \sqrt{N_\mathrm{co}} $).
This is corrected, given that we do not see the theoretical increase in sensitivity, in the following Section.

\subsubsection{Flux density estimation}\label{sec:mwa_flux_density}
The output of the coherent beamforming pipeline (see Section~\ref{sec:mwa_coherent_bforming}) is a set of PSRFITS files \citep{2004PASAHotan}, one file per 200 seconds per $1.28\mathrm{\,MHz}$ coarse channel.
The individual channels can be combined into one 200 second file, reducing the number of data files by a factor of 24. 
These PSRFITS data were then incoherently dedispersed and subdivided into single-pulse archives using \textsc{dspsr} and the ephemeris derived from the Parkes $ 732\mathrm{\,MHz} $ data.
Each coarse channel's edges were flagged (fine channels 0--19 and 108--127) to mitigate the effects of aliasing introduced during the channelization process, and the \textsc{psrchive} routine \textsc{paz} was used to apply the inbuilt median smoothed difference excision algorithm.
Finally, the archives were collapsed in polarization and frequency and written to a time series using \textsc{pdv}, without automatic baseline removal.

To compensate for the fact that the beam simulations assume the ideal $ \sqrt{N_\mathrm{co}} $ improvement in sensitivity, we estimate a coherency factor, $ f_c $, by evaluating 
\begin{equation}\label{eq:coherency}
f_c = \sqrt{N_\mathrm{co}}\left(\frac{\mathrm{(S/N)_{co}}}{\mathrm{(S/N)_{inco}}}\right)^{-1},
\end{equation}
where $\mathrm{(S/N)_{co}}$ is the signal-to-noise ratio of a bright pulse in the coherently beamformed data and $\mathrm{(S/N)_{inco}}$ is the signal-to-noise ratio of the same pulse in the incoherently summed data.
This quantity defines how well the coherent beamforming pipeline performed compared to the theoretical expectation.
The system temperature and gain calculations are used to convert the time series data from arbitrary power units to flux density units using 
\begin{equation}
S = (\mathrm{S/N})\times \frac{f_c T_\mathrm{sys}}{G\sqrt{n \Delta\nu \Delta t}},
\label{eq:radiometer}
\end{equation}
where $ \mathrm{S/N} $ is the sample signal-to-noise ratio, $ f_c $ is the coherency factor as in equation~\eqref{eq:coherency}, $ n $ is the number of polarizations summed (in this case $ n=2 $), $ \Delta\nu $ is the observing bandwidth, and $ \Delta t $ is the sample integration time. 
For an individual MWA tile, the SEFD is typically $ \sim 2\times 10^4 $\,Jy , however, for the coherently beamformed we find (for this set of subbands and pointings) the SEFD to be $ \sim 2\text{--}3\times10^3 $\,Jy.

\section{ANALYSIS AND RESULTS}\label{sec:results}
After post-processing, we produced five time series with $ \Delta t=261.241\mathrm{\,\mu s} $ time resolution.
This was achieved by re-binning the data into 129 phase bins per pulse period, ensuring that both the MWA and Parkes data had a sample time greater than the Parkes intrinsic sampling time $ 256\mathrm{\,\mu s} $.
We use fluence (integrated flux density over the pulse width) as a direct measure of the pulse energy, given that peak or mean flux densities are less informative at MWA frequencies where giant pulses are typically scattered over several pulse periods.

\subsection{Detecting giant pulses}\label{sec:detecting_GPs}
\begin{figure}
\includegraphics[width=\linewidth]{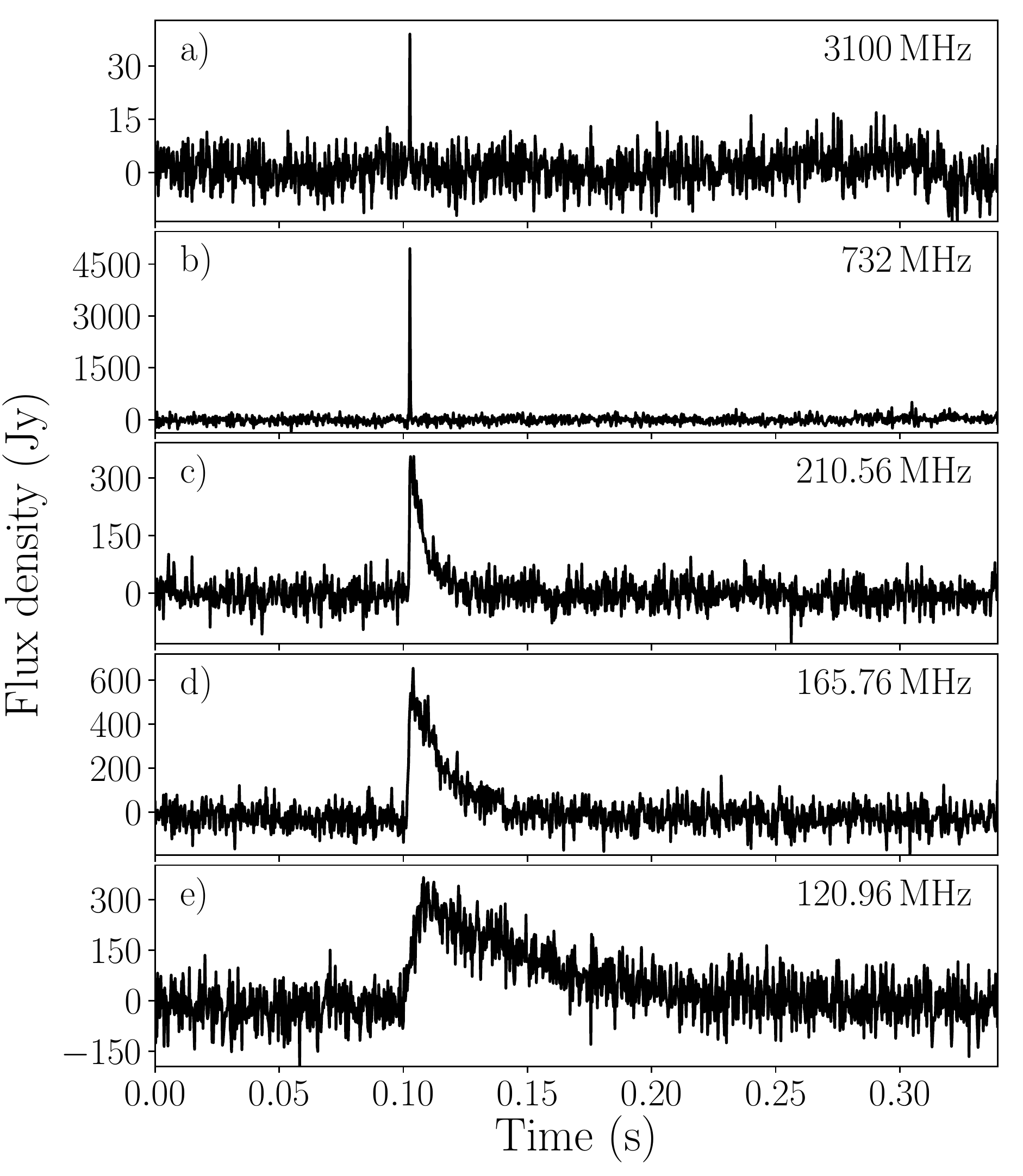}
\caption{A simultaneous giant pulse detected in all five observing bands: a) $ 3100\mathrm{\,MHz} $, b) $ 732\mathrm{\,MHz} $, c) $ 210.56\mathrm{\,MHz} $, d) $ 165.76\mathrm{\,MHz} $, and e) $ 120.96\mathrm{\,MHz} $. 
The effect of multipath scattering are most obvious at MWA frequencies, introducing a significant exponential tail to each giant pulse, while the Parkes pulses are delta-functions with the recorded time resolution. 
The 120.96\,MHz pulse also has a visible rise time compared to the other frequencies.}
\label{fig:aligned_pulse}
\end{figure}

Due to frequency dependent propagation effects, the Parkes and MWA data were processed differently.
As the data were incoherently dedispersed, the dispersive smearing across individual channels was not removed.
While this delay is large at MWA frequencies ($ \sim 1\text{--}10\% $ of a pulse period), the dominating factor is still the multipath scattering which broadens an individual giant pulse across several pulse periods (see Section~\ref{sec:scattering}).
Not only does this scattering make pulse detection and cross-matching more difficult, it also requires a more complicated method of measuring the pulse fluences.
An example of a giant pulse detected simultaneously across all five subbands, shown in Figure~\ref{fig:aligned_pulse}, illustrates the pulse-shape evolution with frequency due to multipath scattering.

A summary of the detected main pulse (MP) and interpulse (IP) giant pulses from each frequency band is presented in Table~\ref{tab:numGPs}.
Every giant pulse detected is recorded in a table format including the pulse number, phase position, and fluence estimate.

\begin{table}[!htbp]
\centering
\caption{Number of detected giant pulses per frequency.\label{tab:numGPs}}
\begin{tabular}{cc|cc}
\hline 
Centre frequency & $N_\mathrm{total}$ & $ N_\mathrm{MP} $ & $ N_\mathrm{IP} $ \\
(MHz) & & & \\
\hline\hline
120.96 & \nMWAlo & \nMWAloMP & \nMWAloIP \\
165.76 & \nMWAmi & \nMWAmiMP & \nMWAmiIP \\
184.96 & \nMWAcont & \nMWAcontMP & \nMWAcontIP \\
210.56 & \nMWAhi & \nMWAhiMP & \nMWAhiIP \\
732    & \nPKSlo & \nPKSloMP & \nPKSloIP \\
3100   & \nPKShi & \nPKShiMP & \nPKShiIP \\
\hline
\end{tabular}
\end{table}

\subsubsection{Parkes}\label{sec:pks_GPs}
The calibrated single-pulse archives for both the 732 and 3100\,MHz data were summed in polarization and frequency to produce total intensity profiles. 
To find giant pulses in the single-pulse archives we used \textsc{psrchive}'s single-pulse analysis routine \textsc{psrspa} to search for candidate events with a signal-to-noise ratio $ \mathrm{SNR}\geq 6 $.
The candidate lists were filtered to remove events with large pulse widths\footnote{The Parkes data is limited by the time resolution, thus giant pulses appear as events with a width of 1 sample only.}.
The time-of-arrival (TOA) was calculated for each giant pulse candidate with the ephemeris used during the folding process.
The giant pulse positions in rotation phase were then examined using \textsc{tempo2}.
At 732 and 3100\,MHz, there were 179 and 39 outliers (main pulse and interpulse combined) discarded, respectively.

This produced a list of \nPKShi\ pulses at 3100\,MHz and \nPKSlo\ pulses at 732\,MHz.
From the finalized list of candidates, on-pulse peak flux densities were recorded for each single-pulse archive.
The giant pulse fluences were then calculated as the product of the peak flux density and the time series bin width.
The fluence errors were calculated from the off-pulse root-mean-square (RMS) value.

Assuming Gaussian noise, the probability, $ P_n $, of a false detection above some signal-to-noise ratio $ n\sigma $ is,
\begin{equation}
P_n(x>n\sigma) = \int_{\mu+n\sigma}^{\infty}P(x)\,\mathrm{d}x = \frac{1}{2}\mathrm{erfc}\left(\frac{n}{\sqrt{2}}\right),
\label{eq:false_detect_prob}
\end{equation}
where $ \mu $ is the mean noise level, $ \sigma $ is the root-mean-square noise, and $ \mathrm{erfc}(x) $ is the complementary error function. 
The signal-to-noise ratio threshold when searching for single pulses in both the 732 and $ 3100\mathrm{\,MHz} $ bands was $ 6\sigma $, which corresponds to a false detection likelihood of $ P_6 \approx 1\times 10^{-9} $.  
The number of false positives ($ N_\mathrm{f,pks} $) is then the product of $ P_6 $ and the number of observed pulsar rotations ($ \approx 1.5\times 10^5 $).
We calculate this number to be significantly less than unity ($ N_\mathrm{f,pks}\approx 1.5\times 10^{-4} $) and therefore do not expect any giant pulse candidates with $ \mathrm{SNR}\geq 6 $ to be spurious.
After removing the RFI, ensuring pulses were recorded only if they occur in the main pulse and interpulse phase windows and by excluding candidates with pulse widths greater than 1 sample, we assert that all Parkes giant pulse candidates used in the following analysis are real.

\subsubsection{MWA}\label{sec:mwa_GPs}
As MWA giant pulses are severely scattered, some custom software was developed specifically for searching for scattered pulses in the time series.
The input to this code is the time series created in Section~\ref{sec:mwa_obs}.
For each time series, the data were smoothed using a Savitzky-Golay low-pass filter with a window length of 9 samples and a 3rd order fitting polynomial. 
Typical pulse widths are $ >40 $ samples, thus the smoothing window length will not adversely affect the local pulse shape.
This mitigated the high-frequency noise without reducing the fidelity of the individual pulses.
The baselines for each time series were then removed by subtracting a linear fit over adjacent $ 10^4 $ sample windows.

Local peaks were detected above a threshold of $ 5.5\sigma $\footnote{A lower threshold than that used for the Parkes observations is enforced for the MWA data because of the significantly quieter RFI environment at the Murchison Radio-astronomy Observatory.} in partially overlapping sections of the time series.
Any new candidate peaks recorded with the same sample number as a previously detected peak were discarded.
Around each of the peaks, between 500 and 1000 samples (from the highest to the lowest frequency, respectively) were retrieved before and after the peak to ensure the entire scattered pulse is captured in the time series window.

In order to further constrain the pulse position and extent, we fitted a pulse broadening function (PBF) to each time series windows.
The thick, finite extent scattering screen PBF proposed by \citet{1972MNRASWilliamson} was chosen, as it models both the significant rise-time and exponential scattering tail present at low frequencies.
The sample selections were fitted with the corresponding PBF form,
\begin{equation}
g(t) = A\left(\frac{\pi\tau_\mathrm{d}}{4\left(t-t_0\right)^3}\right)^{1/2}\exp\left[-\frac{\pi^2\tau_\mathrm{d}}{16\left(t-t_0\right)}\right]
\label{eq:pbf}
\end{equation}
where $ A $ is a constant amplitude scaling, $ t_0 $ is the start time of the leading edge of the pulse and $ \tau_\mathrm{d} $ is the characteristic scattering time.
Pulse numbers and phase were calculated based on the best-fitting pulse starting time, $t_0$.
The pulse candidates were then selected based on whether their fitted $ \tau_\mathrm{d} $ values fell within a predetermined range, based on the approximate scattering time measured at each frequency (see Section~\ref{sec:scattering}).
This distinguishes bona fide candidates with sensible scattering time estimates from spurious detections.
We note that the fitting was used only as a filtering process, and because we know that, for the Crab, none of the standard PBFs fit correctly (see Kirsten et al. in prep.) the resulting scattering times may be somewhat less reliable.
Each fitted pulse was also inspected by eye so that any questionable candidates were removed.
For the full-bandwidth observation (184.96\,MHz), this produced a list of \nMWAcont\ pulses.
For the split-bandwidth observation, we record: \nMWAlo\ pulses at 120.96\,MHz; \nMWAmi\ pulses at 165.76\,MHz; and \nMWAhi\ pulses at 210.56\,MHz. 

For each real candidate pulse, we define the start of the pulse as the best-fitted $ t_0 $, and the end of the pulse as 6 e-folds past the PBF peak (i.e. $ t_0+ \pi^2\tau_\mathrm{d}/4 + 6\hat{\tau}_\mathrm{d}$). 
In this case, $ \hat{\tau}_\mathrm{d} $ is the median scattering timescale as in Table~\ref{tab:GPscattering} while $\tau_\mathrm{d}$ is the best-fitting scattering time for the individual pulse.
We define this window as the actual pulse from which to calculate the fluence.
For each candidate we then integrate over the pulse window and record that as the pulse fluence, along with the fluence from the fitted PBF.
The fluence uncertainty for each pulse was calculated by integrating under the fitted PBF model, scaled such that the peak amplitude was equal to the local RMS value.

Detections near the threshold limit may have underestimated fluences, given that the giant pulses (specifically the scattered tail) would be dominated by noise and fall within the baseline RMS well before a brighter counterpart pulse at a different frequency.
Fluence estimates, and consequently the calculated spectral indices (see Section~\ref{sec:gp_spectra}), in those cases may be less reliable, especially for weaker pulses.  
Additionally, the software searches only for simple PBF forms, thus giant pulses with significantly different structure (e.g. a second pulse within the scattering tail) may be discarded, especially if the structure is such that the estimated scattering time scales are outside the nominally expected range.
At 210.56, 184.96, 165.76 and 120.96\,MHz, this results in $\sim 7\%$, $\sim 0.4\%$, $ \sim 10\% $ and $ \sim 1\% $ of candidates being flagged, respectively. 
The 184.96\,MHz fraction is significantly smaller due to both the sensitivity (i.e. the noise characteristics are typically better behaved) and the observation duration (i.e. we are less likely to observe, for instance, a giant pulse within the scattering tail of another).

\begin{figure}
\includegraphics[width=\linewidth]{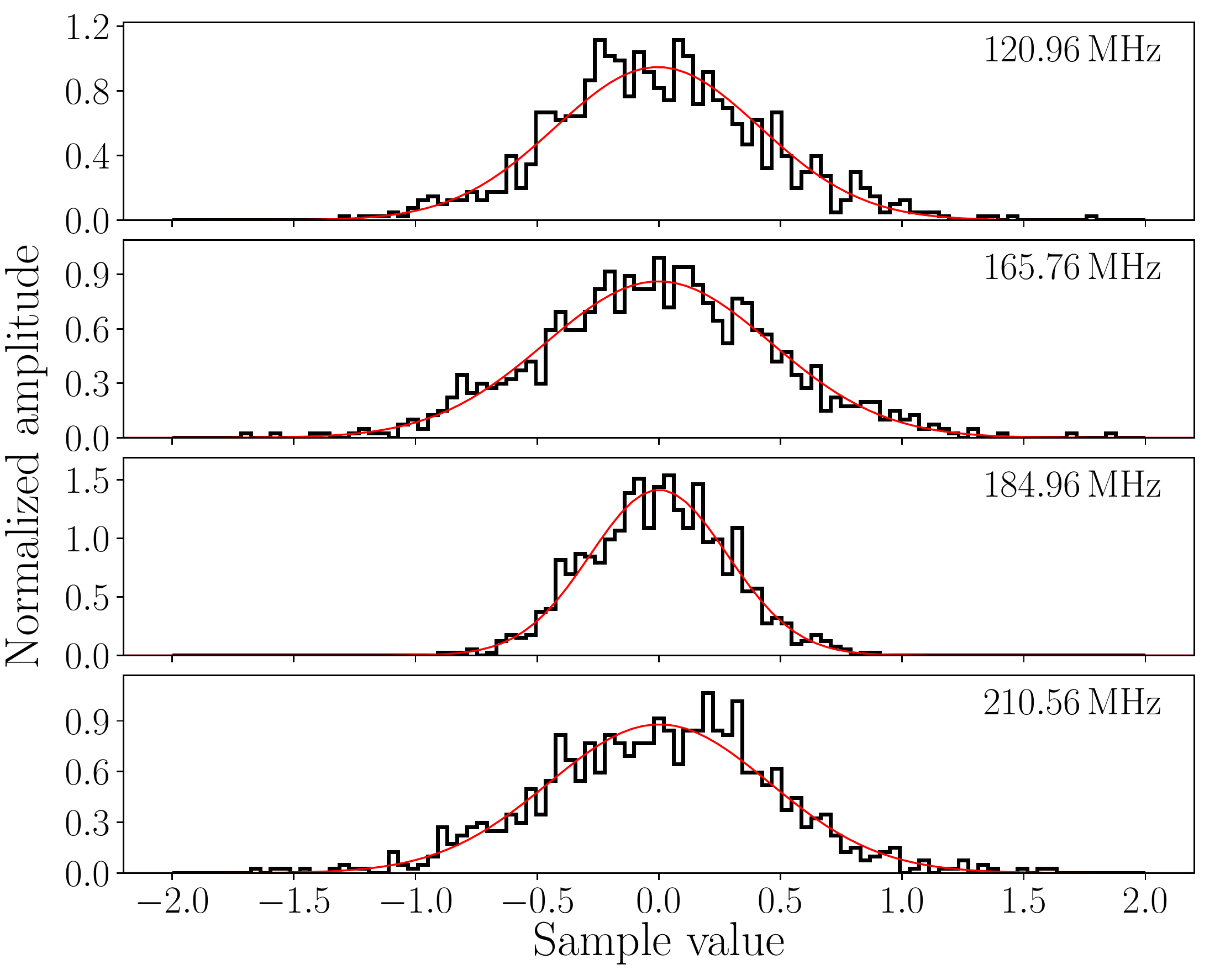}
\caption{A 1000 sample example of the noise characteristics for coherently beamformed, dedispersed, baseline-removed MWA data.
The red solid line is a fitted normal probability density function.}
\label{fig:coherent_bf_noise}
\end{figure}

We tested the noise statistics for coherently beamformed, dedispersed, baseline-removed MWA data for normality.
This was achieved by selecting five evenly spaced samples, each containing 1000 data points, from each subband time series and fitting a normal distribution.
From these samples, the noise statistics are consistent with Gaussian noise (see Figure~\ref{fig:coherent_bf_noise} for an example), therefore we can use equation~\eqref{eq:false_detect_prob} to calculate the false detection likelihood for MWA data.
The signal-to-noise ratio threshold when searching through MWA data was $ 5.5\sigma $, thus the false detection probability is $ P_{5.5} \approx 2\times 10^{-8} $.
The number of pulsar rotations during the MWA split-bandwidth observations is $ \approx 1.1\times 10^5 $, thus the number of false detections expected is again significantly less than unity ($ N_\mathrm{f,mwa,split}\approx 2.2\times 10^{-3} $).
For the full-bandwidth observation, the number of pulsar rotations is $ \approx 3.5\times 10^4 $, and the number of expected false detections is $ N_\mathrm{f,mwa,full}\approx 7\times 10^{-4} $ -- again much less than unity.
We claim that no MWA giant pulses are spurious detections, given the statistics and the filtering performed during the candidate selection process.

\subsection{Pulse broadening}\label{sec:scattering}
\begin{figure}[htpb]
\includegraphics[width=\linewidth]{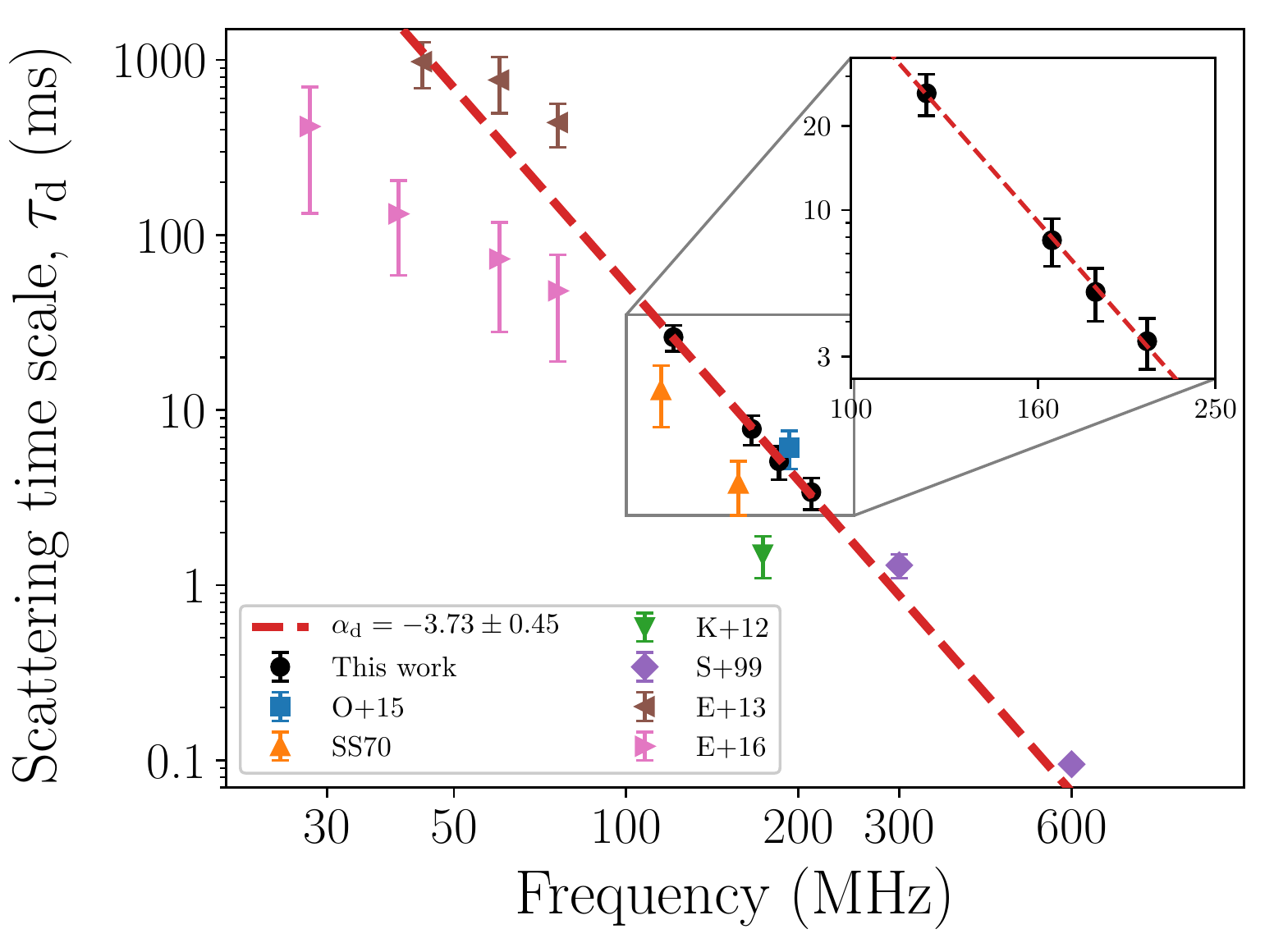}
\caption{Pulse broadening times from the four MWA bands.
The median scattering time scales (circles), their respective errors and the fitted power-law (red dashed line), are plotted on a log-log scale.
The scaling index, $ \alpha_\mathrm{d}=-3.73\pm0.45 $ is significantly shallower than what is expected from a Kolmogorov model, but consistent with other estimates at similar frequencies in the literature.
Given the variability of Crab pulse broadening, it is not surprising that many of the scattering times from the literature do not fall on the fitted power-law. 
The scattering times legend keys are as in Table~\ref{tab:GPscattering}.}
\label{fig:scattering}
\end{figure}

At both Parkes subbands, we cannot directly determine the scattering time scale ($ \tau_\mathrm{d} $) since we are limited by the time resolution (261.241\,$\mu$s) of our recorded data.
At MWA frequencies, from the rudimentary fitting performed when detecting the pulses we can estimate the pulse broadening.
We report the median scattering time scales in Table~\ref{tab:GPscattering}.
Furthermore, we calculated the scattering spectral index ($\alpha_\mathrm{d}$) using the MWA data.
Using a least-squares minimization approach, we fitted a power-law ($ \tau_\mathrm{d}\propto \nu^{\alpha_\mathrm{d}} $) to the MWA scattering time scales (see Figure~\ref{fig:scattering}).
The determined scaling index is $ \alpha_\mathrm{d}=-3.73\pm0.45 $, significantly shallower than what is predicted from a Kolmogorov model, which is $ \alpha_\mathrm{d}=-4.4 $.
This results is consistent with what is reported in the literature at low frequencies (e.g. \citealp{2007ApJBhat,2013ApJEllingson,2016ApJEftekhari}).
Extrapolating using the above scaling index, we also estimate the scattering expected in the Parkes subbands in Table~\ref{tab:GPscattering}.

\begin{table*}[!htbp]
\centering
\caption{Pulse broadening time scales in the literature, including this work.\label{tab:GPscattering}}
\begin{tabular}{cccc}
\hline 
Centre frequency & $\tau_\mathrm{d}$ & Reference & Key\\
(MHz) & (ms) & &\\
\hline\hline
28 & $417\pm284$ & \citet{2016ApJEftekhari} & E+16 \\
40 & $132\pm73$ & \citet{2016ApJEftekhari} & E+16 \\
44 & $978\pm287$ & \citet{2013ApJEllingson} & E+13 \\
60 & $768\pm273$ & \citet{2013ApJEllingson} & E+13 \\
60 & $73\pm45$ & \citet{2016ApJEftekhari} & E+16 \\
76 & $48\pm29$ & \citet{2016ApJEftekhari} & E+16 \\
76 & $439\pm122$ & \citet{2013ApJEllingson} & E+13 \\
115 & $13\pm5$ & \citet{1970NatStaelin} & SS70 \\
120.96 & $26.1\pm4.4$ & \textit{This work} & \\
157 & $3.8\pm1.3$ & \citet{1970NatStaelin} & SS70 \\
165.76 & $7.8\pm1.5$ & \textit{This work} &\\
173.25\tablenotemark{$a$} & $1.5\pm0.4$ & \citet{2012AAKaruppusamy} & K+12\\
184.96 & $5.1\pm1.1$ & \textit{This work} &\\
192.64 & $6.1\pm1.5$ & \citet{2015ApJOronsaye} & O+15\\
210.56 & $3.4\pm0.7$ & \textit{This work} &\\
300 & $1.3\pm0.2$ & \citet{1999ApJSallmen} & S+99\\
600 & $0.095\pm0.005$ & \citet{1999ApJSallmen} & S+99\\
732    & $ \sim0.03 $\tablenotemark{$b$} & \textit{This work} &\\
3100   & $ \sim0.0002 $\tablenotemark{$b$} & \textit{This work} &\\
\hline
\end{tabular}
\tablenotetext{a}{Scattering time estimated from Figure 6 in \citet{2012AAKaruppusamy}.}
\tablenotetext{b}{Extrapolated from 184.96\,MHz, assuming $ \tau_\mathrm{d}\propto\nu^{-3.7} $.}
\end{table*}

Given the time-variability of characteristic scattering times observed for the Crab, and the dependence of the estimated scattering time on the chosen PBF, discrepancies as much as by a factor of $ \sim 2 $ are not uncommon between similar frequencies.
Our values from the MWA subbands are roughly consistent with those quoted in the literature (e.g. \citealp{1970NatStaelin,2006ARepPopov,2015ApJOronsaye}).
A more detailed examination of the scattering behavior of the Crab and other pulsars within the MWA observing frequency range will be reported in a forthcoming publication (Kirsten et al., in prep.).
In particular, there is discussion of the difficulties in correctly characterizing the pulse broadening seen in Crab giant pulses at low frequencies and reconciling this with a variety of theoretical scattering screen models.

\subsection{Simultaneous giant pulses}\label{sec:sim_GPs}
For every giant pulse found in Sections~\ref{sec:pks_GPs} and \ref{sec:mwa_GPs}, a pulse number was recorded.
We use those pulse numbers and the phase (to discriminate between MP and IP giant pulses) of each giant pulse to cross-match across the five frequency bands.
The cross-matching was achieved by using routines from the Starlink Tables Infrastructure Library Tool Set (STILTS; \citealp{2006ASPCTaylor}), which is designed for robust and efficient processing of tabular data.
The tools are implemented for generic manipulation of tabulated data sets, though are typically used for astronomical object catalog analysis, in particular cross-matching of large data sets based on user-specified selection criteria.
The results of cross-matching the giant pulse samples from each subband are summarized in Table~\ref{tab:simul_GPs}.

\begin{table*}
\renewcommand{\arraystretch}{1.3}
\renewcommand{\arrayrulewidth}{1.5pt}
\centering
\caption{Number of simultaneous pulses between different frequency bands. 
The table is split by the diagonal (left-to-right): numbers in white cells represent simultaneous main pulses, while numbers in gray cells represent simultaneous interpulses. 
The values along the diagonal (in light blue), separated by a backslash (\textbackslash) are the number of main pulses (left) and interpulses (right) for each band. 
Columns marked with a dash (--) indicates no cross-matching was possible. \label{tab:simul_GPs}}
\begin{tabular}{c|CCCCCC}
Frequency (MHz) & 120.96 & 165.76 & 184.96 & 210.56 & 732 & 3100 \\
\hline
\RowColor 120.96 & \allc\nMWAloMP\,\textbackslash\,\nMWAloIP & 8  & -- & 7  & 4  & 1 \\
\RowColor 165.76 & \mpc 72 & \allc\nMWAmiMP\,\textbackslash\,\nMWAmiIP & -- & 42 & 25 & 5 \\
\RowColor 184.96 & \mpc -- & \mpc --  & \allc\nMWAcontMP\,\textbackslash\,\nMWAcontIP & -- & 33 & 2 \\
\RowColor 210.56 & \mpc 66 & \mpc 269 & \mpc --  & \allc\nMWAhiMP\,\textbackslash\,\nMWAhiIP  & 56 & 5 \\
\RowColor 732    & \mpc 38 & \mpc 173 & \mpc 141 & \mpc 326 & \allc\nPKSloMP\,\textbackslash\,\nPKSloIP & 9 \\
\RowColor 3100   & \mpc 6  & \mpc 16  & \mpc 10  & \mpc 22  & \mpc 157 & \allc\nPKShiMP\,\textbackslash\,\nPKShiIP \\
\end{tabular}
\end{table*}

Between the two Parkes frequencies, we find that there are 157 simultaneous main pulses and 9 simultaneous interpulses. 
These numbers correspond to approximately $ 72\% $ and $ 64\% $ coincidence for main pulses and interpulses respectively, based on the number of pulses detected in the $ 3100\mathrm{\,MHz} $ band. 

Between the MWA full-bandwidth observation and the 732\,MHz Parkes band, there are 140 simultaneous main pulses and 33 simultaneous interpulses, corresponding to 41\% and 50\% based on the total numbers from the 184.96\,MHz band.
Across all three bands, we detected 10 simultaneous main pulse giant pulses and 2 simultaneous interpulse giant pulses.

Within the MWA bands (Table~\ref{tab:observations}), the full-bandwidth and split-bandwidth observations have no overlap in time, thus we focus only on the three subbands at 120.96, 165.7, and 210.56\,MHz. 
Between the highest and middle bands, there are 269 simultaneous main pulses and 42 interpulses, corresponding to 80\% and 84\% based on the number of pulses detected in the 165.76\,MHz band.
Between the lowest and middle bands there are 68 simultaneous main pulses and 8 interpulses, corresponding to 87\% and 72\% correlation based on the number of pulses detected in the 120.96\,MHz band. 
There are 7 giant pulses detected simultaneously across all five bands, 6 main pulses and 1 interpulse.

For the brightest $ \sim 10\% $ of pulses (combining main pulses and interpulses) in each band, we checked for pulses that had no counterpart in adjacent frequency bands.
At 210.56 MHz, there are 68 pulses with fluences greater than 1.5\,Jy\,s, of which there are only 42 counterparts at 732\,MHz and 67 counterparts at 165.76\,MHz.
Inspecting the MWA time series, we found that the missing giant pulse in the 165.76\,MHz band is below the detection threshold.
At 165.76 MHz, there are 38 pulse with fluences greater than 5\,Jy\,s, with 37 counterparts at 210.56\,MHz and 26 counterparts at 120.96\,MHz.
The missing counterpart at 210.56\,MHz is relatively clear in the time series, however it is actually two giant pulses combined (and therefore discarded during the candidate processing) -- a main pulse and interpulse in adjacent rotations.
At 732\,MHz, the main pulse is detected, but the interpulse in the subsequent rotation is not.
For the 12 missing counterparts at 120.96\,MHz, in 8 of those cases, there is a visible counterpart below the $5.5\sigma$ detection threshold.
For another 3, there are no visible counterparts.
For one pulse, there is no 120.96\,MHz data at the corresponding time because of the dispersion delay.

In light of the ``double giant pulse'' (i.e. a main pulse and interpulse occurring within one rotation), we searched for other examples across all frequency bands.
At 3100\,MHz, there is one marginal case ($ \sim 0.4\% $ of detected pulses), while at 732\,MHz there are 85 clear examples ($ \sim 1\% $ of detected pulses).
Within the MWA bands, the pulse broadening makes robustly identifying double giant pulses difficult, however, we find $ \sim 1\text{--}2 $ marginal examples per MWA band.
The double giant pulses at one band do not necessarily coincide with double giant pulses at any other.

\subsection{Giant pulse fluence distributions}\label{sec:gp_fluences}
\begin{figure}
\includegraphics[width=\linewidth]{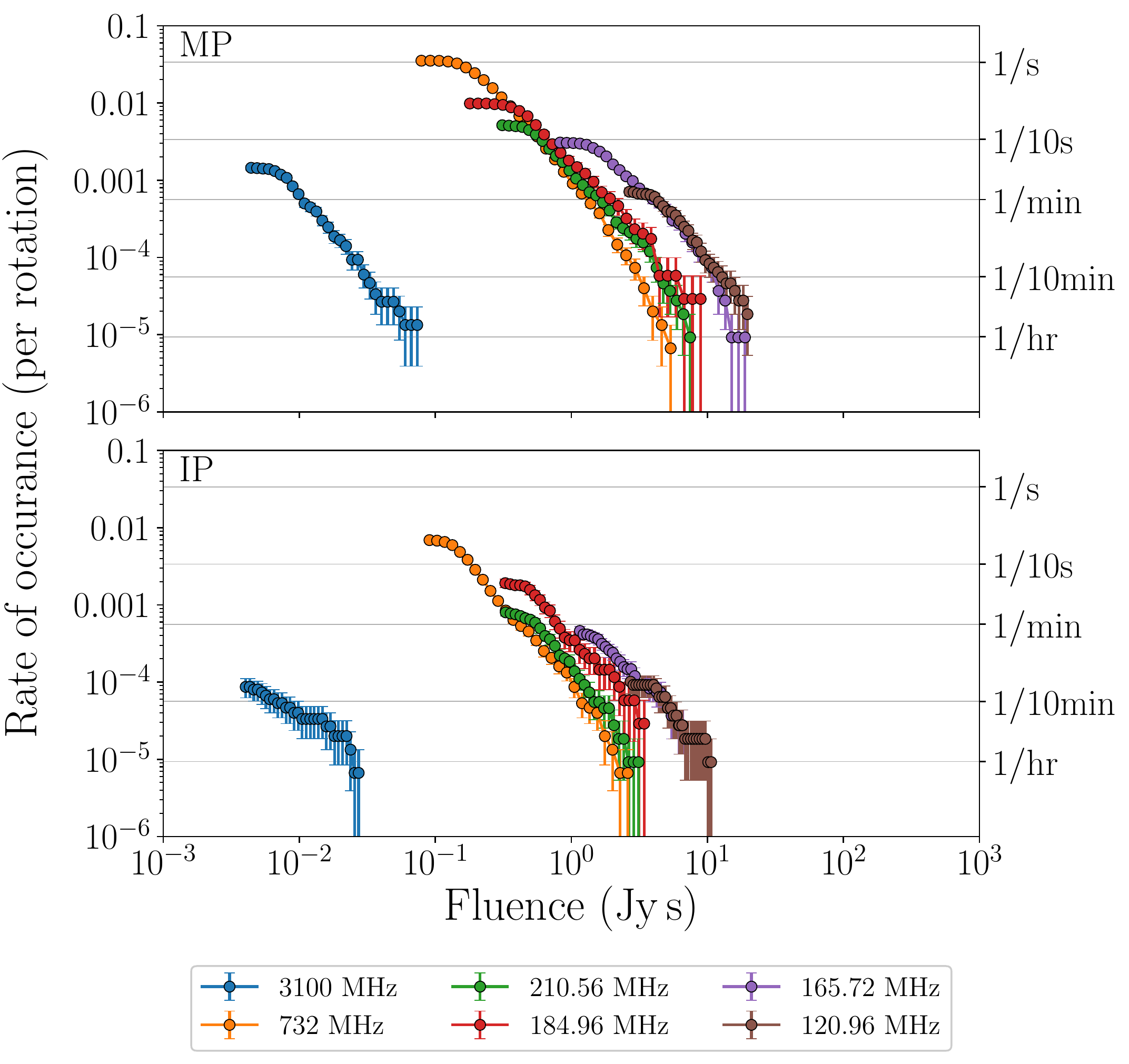}
\caption{Giant pulse rates versus fluence for each observed frequency
The left vertical axis effectively denotes the probability of detecting one giant pulse per rotation, while on the right these are translated into rates.
The clustering of the low-frequency bands (120.96, 165.76, 185.96 and 210.56\,MHz) hints that the spectral index is flattening for both main pulses (top) and interpulses (bottom).}
\label{fig:pulse_rates}
\end{figure}

In Figure~\ref{fig:pulse_rates} we plot the complementary cumulative distribution function (CCDF), also known as the survival function, of pulses as a function of fluence for each subband.
The clustering at low frequencies suggests that there is some degree of flattening of the spectral indices occurring at the lowest frequencies.
This also provides estimates for sub-populations of giant pulses and rates of occurrence as a function of frequency and fluence.
Listed in Table~\ref{tab:GPfluences} are some basic quantities describing the fluences of all detected giant pulses in each observed band.

\begin{table*}
\renewcommand{\arraystretch}{1.2}
\centering
\caption{General statistics of the full sample of giant pulse fluences from each subband.\label{tab:GPfluences}}
\begin{tabular}{ccccccccc}
\hline 
 & \multicolumn{4}{c}{Main pulse $ F_\nu $ } & \multicolumn{4}{c}{Interpulse $ F_\nu $ }\\
\cline{2-9}
Center frequency & Median & Std. dev. & Min. & Max. & Median & Std. dev. & Min. & Max. \\
(MHz) & (Jy\,s) & (Jy\,s) & (Jy\,s) & (Jy\,s) & (Jy\,s) & (Jy\,s) & (Jy\,s) & (Jy\,s)\\
\hline\hline
120.96 & 5.62 & 3.71 & 2.57 & 20.42 & 4.91 & 2.34 & 2.66 & 10.91\\
165.76 & 1.99 & 2.46 & 0.78 & 19.96 & 2.01 & 1.52 & 1.12 & 6.77\\
184.96\tablenotemark{$a$} & 0.52 & 0.78 & 0.16 & 9.54 & 0.61 & 0.58 & 0.31 & 3.57\\
210.56 & 0.64 & 0.80 & 0.29 & 7.89 & 0.61 & 0.54 & 0.31 & 3.53\\ 
732    & 0.22 & 0.28 & 0.07 & 5.77 & 0.17 & 0.22 & 0.08 & 3.58\\
3100   & 0.009 & 0.009 & 0.004 & 0.077 & 0.008 & 0.008 & 0.004 & 0.028\\
\hline
\end{tabular}
\tablenotetext{a}{Adjusting to account for the bandwidth difference produces a median of 1.7\,Jy\,s for main pulses and 1.5\,Jy\,s for interpulses.}
\end{table*}

\begin{figure}
\includegraphics[width=\linewidth]{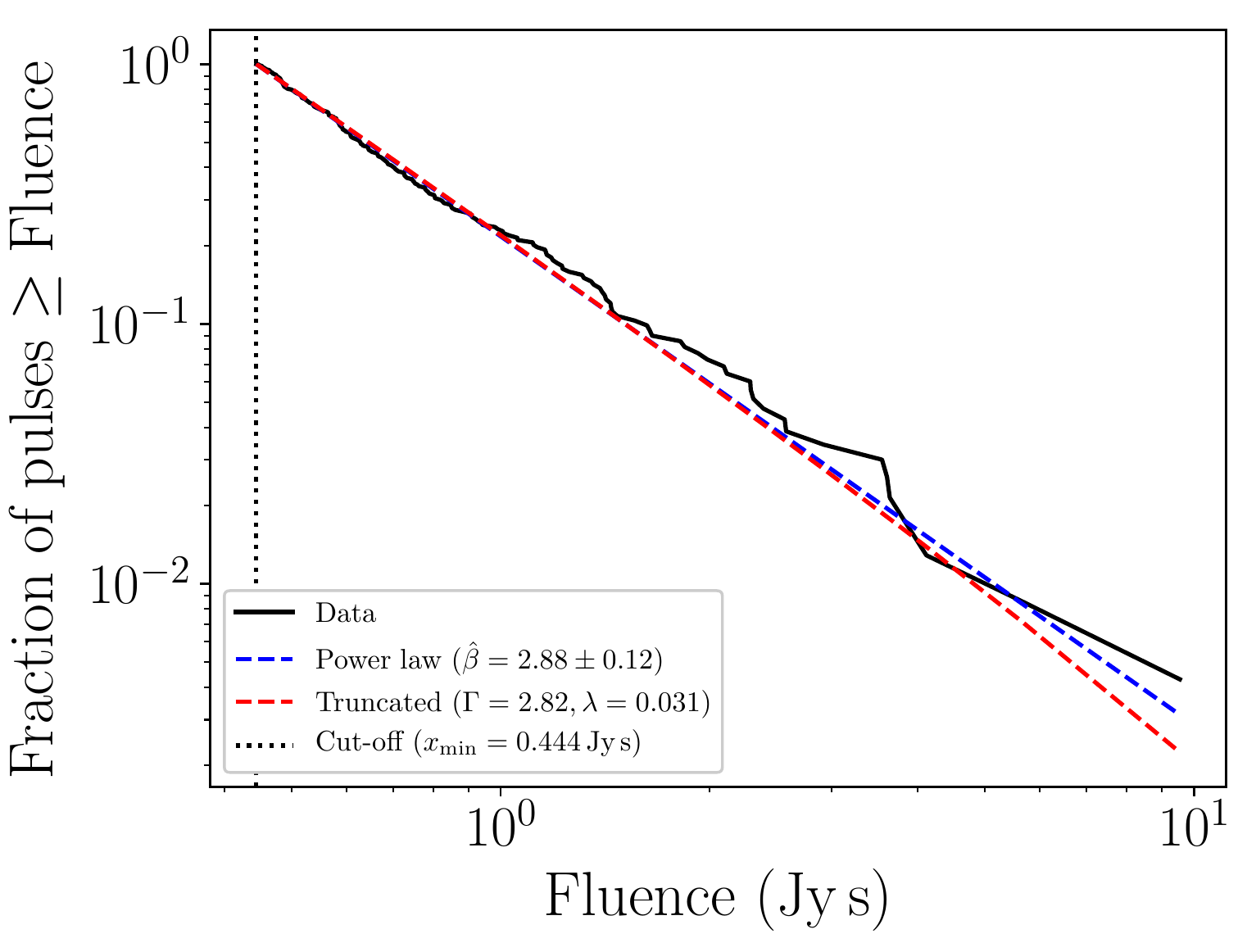}
\caption{The normalized fluence distribution fits for MWA 184.96\,MHz main pulses, where the $y$-axis now represents the fraction of pulses observed greater than a given fluence value up to the power-law cut-off.}
\label{fig:fluence_dist_fits}
\end{figure}

Typically, the fluence distributions are assumed to follow a power-law, $ N(>F_\nu)\propto F_\nu^{-\beta} $.
In the literature, the standard approach is to estimate a power-law cut-off ($ x_\mathrm{min} $, see Figure~\ref{fig:fluence_dist_fits}) by eye and use a least-squares approach to only fit data beyond that limit.
This approach may introduce significant biases in the power-law index estimation and assumes that the data are independent and identically sampled. 

To avoid subjectivity, we chose to use the {\tt powerlaw} Python module \citep{2014PLoSOAlstott}, which appropriately treats several heavy-tailed distributions, particularly focusing on power-laws. 
The best-fitting power-law distribution index ($ \hat{\beta} $) and power-law cut-off ($ x_\mathrm{min} $) are determined by finding the minimum Kolmogorov-Smirnov distance between the data and model (see e.g. Figure~\ref{fig:fluence_dist_fits}).
The data are used to evaluate multiple distribution models, including truncated power-laws (where $ N(>F_\nu)\propto F_\nu^{-\Gamma}e^{-\lambda\nu} $), log-normal, and exponential distributions.
This provides the ability to statistically test which distribution is a better representation of the data based on the likelihood ratios and $ p $-values.
In general, we find that a power-law distribution is the most likely, however the significance of that distinction varies drastically between bands and the compared distributions.
Therefore, we cannot say for certain that a power-law is the best-fitting distribution for all of our data.
Table~\ref{tab:fluence_dist_params} summarizes the fitting results assuming a power-law distribution.

Our results in terms of $ \hat{\beta} $ for the two Parkes subbands are within the range of those reported by \citet{2012ApJMickaliger} between 330 and 1200\,MHz ($ \hat{\beta}_\mathrm{MP}\sim 2.1\text{--}3.1 $, $ \hat{\beta}_\mathrm{IP}\sim 2.4\text{--}3.1 $), but steeper than reported by \citet{2008ApJBhat} between 1300 and 1470\,MHz ($ \hat{\beta} = 2.33\pm0.15$, where the MP and IP are combined), except in the case of our 3100\,MHz Parkes IP exponent.
For the MWA subbands, results are typically steeper than the estimated value at 325\,MHz ($ \hat{\beta}_\mathrm{MP}=2.61\pm0.14 $, $ \hat{\beta}_\mathrm{IP}=2.7\pm0.7 $) reported by \citet{2016ApJMikami}, and are also steeper than the slopes calculated by \citet{2010AAKaruppusamy} between 110--180\,MHz ($ \hat{\beta}_\mathrm{MP}\sim 1.5\text{--}2.4 $, $ \hat{\beta}_\mathrm{IP}\sim 0.7\text{--}2.7 $, with errors typically around $ \pm0.1 $ for main pulses and $ \pm0.5 $ for interpulses).
The MWA main pulse indices are shallower (except for the 120.96\,MHz band) than reported by \citet{2015ApJOronsaye}, where $ \hat{\beta}=3.35\pm0.35 $ (main pulses and interpulses combined), though interpulse indices for all MWA bands are consistent. 

\begin{table*}[!htbp]
\renewcommand{\arraystretch}{1.2}
\centering
\caption{Best-fit parameters for the fluence distributions in each band. \label{tab:fluence_dist_params}}
\begin{tabular}{ccccccc}
\hline
Frequency & \multicolumn{3}{c}{Main pulse} & \multicolumn{3}{c}{Interpulse}\\
\cline{2-7}
(MHz) & $ \hat{\beta} $\tablenotemark{$a$} & $ x_\mathrm{min} $ & $ N>x_\mathrm{min} $ & $ \hat{\beta} $\tablenotemark{$a$} & $ x_\mathrm{min} $ & $ N>x_\mathrm{min} $ \\
\hline\hline
120.96 & $ 3.73\pm0.54 $ & 6.81 & 24  & $ 3.70\pm0.86 $ & 3.94 & 10 \\
165.76 & $ 2.69\pm0.11 $ & 1.60 & 242 & $ 2.84\pm0.29 $ & 1.47 & 40 \\
184.96 & $ 2.88\pm0.12 $ & 0.44 & 234 & $ 3.10\pm0.29 $ & 0.49 & 52 \\
210.56 & $ 2.90\pm0.09 $ & 0.51 & 434 & $ 3.14\pm0.25 $ & 0.49 & 71 \\
732\tablenotemark{$b$}    & $ 3.30\pm0.09 $ & 0.46 & 719 & $ 3.16\pm0.09 $ & 0.15 & 658 \\
3100   & $ 3.19\pm0.17 $ & 0.01 & 82 & $ 2.15\pm0.31 $ & 0.004 & 13 \\
\hline
\end{tabular}
\tablenotetext{a}{The uncertainties quoted are the standard error in the power-law index estimation.}
\tablenotetext{b}{In this case, the evaluated $ x_\mathrm{min} $ for the main pulses is relatively high, excluding $ \sim85\% $ of detected pulses. See text for details.}
\end{table*}

For main pulses at 732, 210.56, and 165.76\,MHz, the distribution appears more likely to be log-normal or a truncated power-law, and the significance ($ p>0.05 $) is such that we cannot entirely reject that hypothesis.
In each of the three bands, only a handful of pulses (i.e. less than 10) contribute to the non-standard power-law shape.
Notably, the determined power-law cut-off for the 732\,MHz data is relatively high compared to the other bands, such that only $\sim15\%$ of pulses are being fit.
If we set an upper limit of 1\,Jy\,s (of which only 2\% of main pulses are brighter) then the re-evaluated power-law fit is such that $ \hat{\beta}=3.12\pm0.04 $ and $ x_\mathrm{min}=0.21$\,Jy\,s and $ >53\% $ of main pulses are included in the fitted distribution.

\subsection{Spectral index distributions}\label{sec:gp_spectra}
\begin{figure*}
\includegraphics[width=\linewidth]{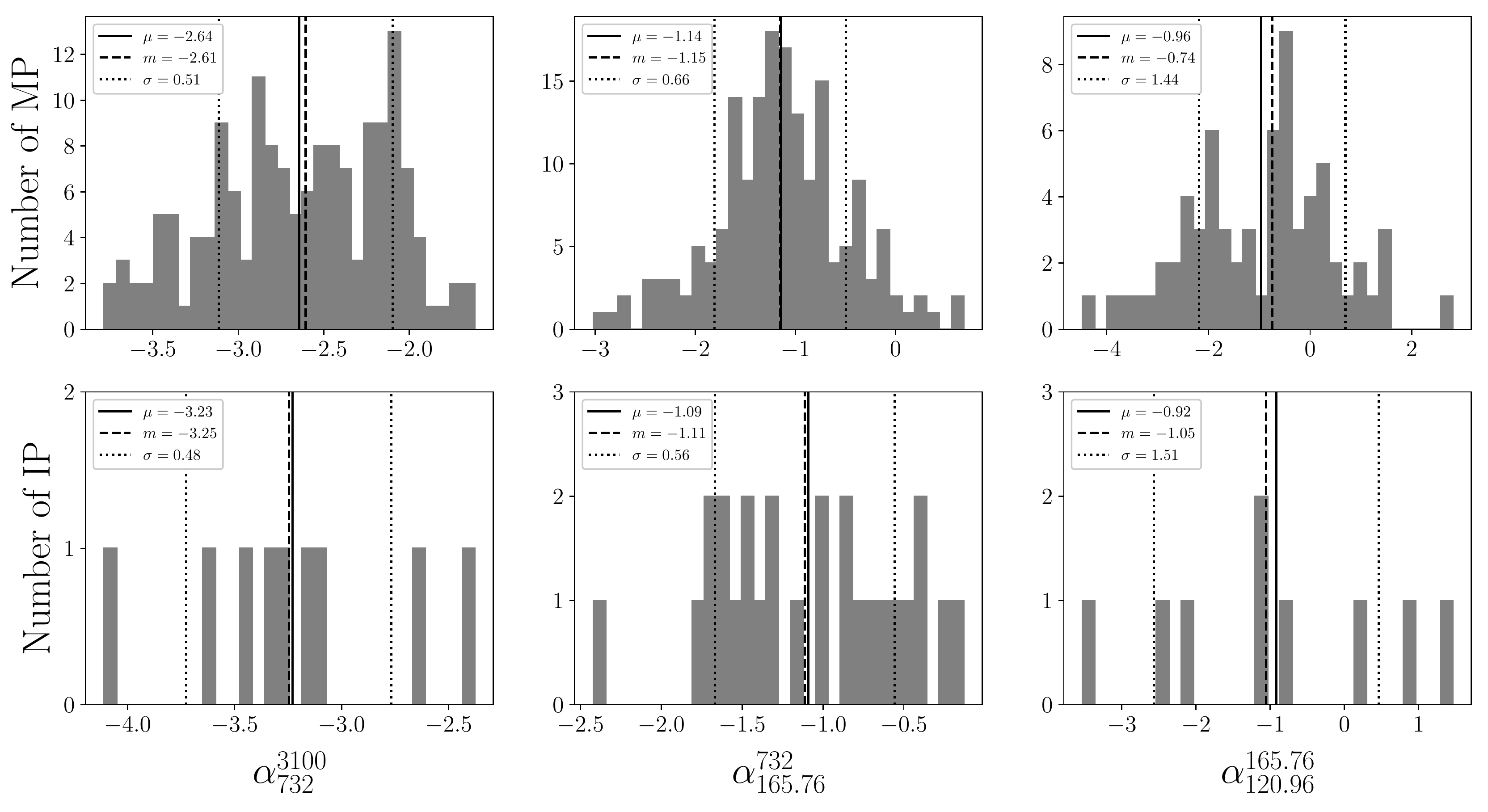}
\caption{Distributions of spectral indices for each combination of giant pulses with the mean ($ \mu $), median ($ m $) and standard deviation ($ \sigma $) included. 
The top row contains only the main pulse simultaneous matches and the bottom row contains only the interpulse matches. 
The spectrum of the simultaneous giant pulses appears to be flattening at low frequencies
There are only a handful of interpulses, thus the mean and median estimates are not as meaningful, but tend to be steeper between the Parkes bands and similar to the main pulses at the lower frequencies.
}
\label{fig:alpha_dists}
\end{figure*}

The spectral index for giant pulse emission is typically assumed to be a power-law, where $ S_\nu\propto \nu^\alpha $, which will also apply to fluences, such that $ F_\nu\propto \nu^\alpha $.
We find that a simple power-law is unable to accurately model the observed giant pulse spectrum between 120.96 and 3100\,MHz.
In Figure~\ref{fig:alpha_dists} we plot the spectral index distributions between each consecutive frequency pair, separated into main pulses and interpulses.
From 732--3100\,MHz, 75\% of simultaneous main pulses have a spectral index between $ -3.3 $ and $ -2.1 $. 
Between 732 and 165.76\,MHz, the same fraction of the giant pulses exhibit a spectral index in the range $ -1.8 $ to $ -0.4 $.
The distribution between the two lowest MWA bands is wider and flatter, with 75\% of pulses within $ -2.5 $ to $ 0.7 $.
Using the 184.96\,MHz data, we also calculated the spectral index distribution for a similar sample of giant pulses (with a signal-to-noise ratio $ \geq 11 $ which accounts for the factor of 2 sensitivity improvement provided by 4 times the bandwidth).
This produces a distribution with a mean $ \alpha=-0.8 $ and a width of $ 0.6 $, with 75\% of the pulses between $ -1.5 $ and $ -0.1 $.
Given the sparse interpulse distributions, we did not calculate the above intervals, though we can say that they appear to follow a similar trend of flattening.

As discussed in Section~\ref{sec:mwa_calibration}, the trustworthiness of the 210.56\,MHz beam, and hence the fluence estimates, are questionable.
We calculate a spectral index from the data in Table~\ref{tab:GPfluences} between 210.56 and 732\,MHz to be $ \alpha\approx-0.6 $ with a width of $ 0.5 $, while between 210.56 and 165.76\,MHz $ \alpha\approx-4.7 $ with a distribution width of $ \sim 3 $.
The 210.56\,MHz data is therefore not used in the following analysis.

\citet{2010AAKaruppusamy} report spectral index distributions between 1300--1450\,MHz centered around $ -1.44\pm3.3 $ and $ -0.6\pm3.5 $ for main pulse and interpulse giant pulses respectively, though the distribution width ranges from approximately $ -15 $ to $ +10 $.
\citet{2016ApJMikami} also estimate spectral indices in the range $ -15 $ to $ +10 $ based on their fluence calculations between 1586--1696\,MHz.
We therefore do not find it surprising that our spectral index distributions are relatively wide, especially between MWA subbands.

Our observations indicate that the spectral index for simultaneous giant pulses is flattening over the sampled frequency range.
If we use the median fluences from Table~\ref{tab:GPfluences}, the computed main pulse spectral index is $ \approx -1.4 $ across most bands, except for between 120.96 and 165.76\,MHz where it steepens to $ \approx -3.3 $ and between the Parkes bands.
In part this is due to the smaller lever-arm available between MWA bands, however it also indicates that the detected simultaneous pulses (which have a slightly shallower spectral index) are more consistent tracers of the spectral flattening.

\begin{figure}
\includegraphics[width=\linewidth]{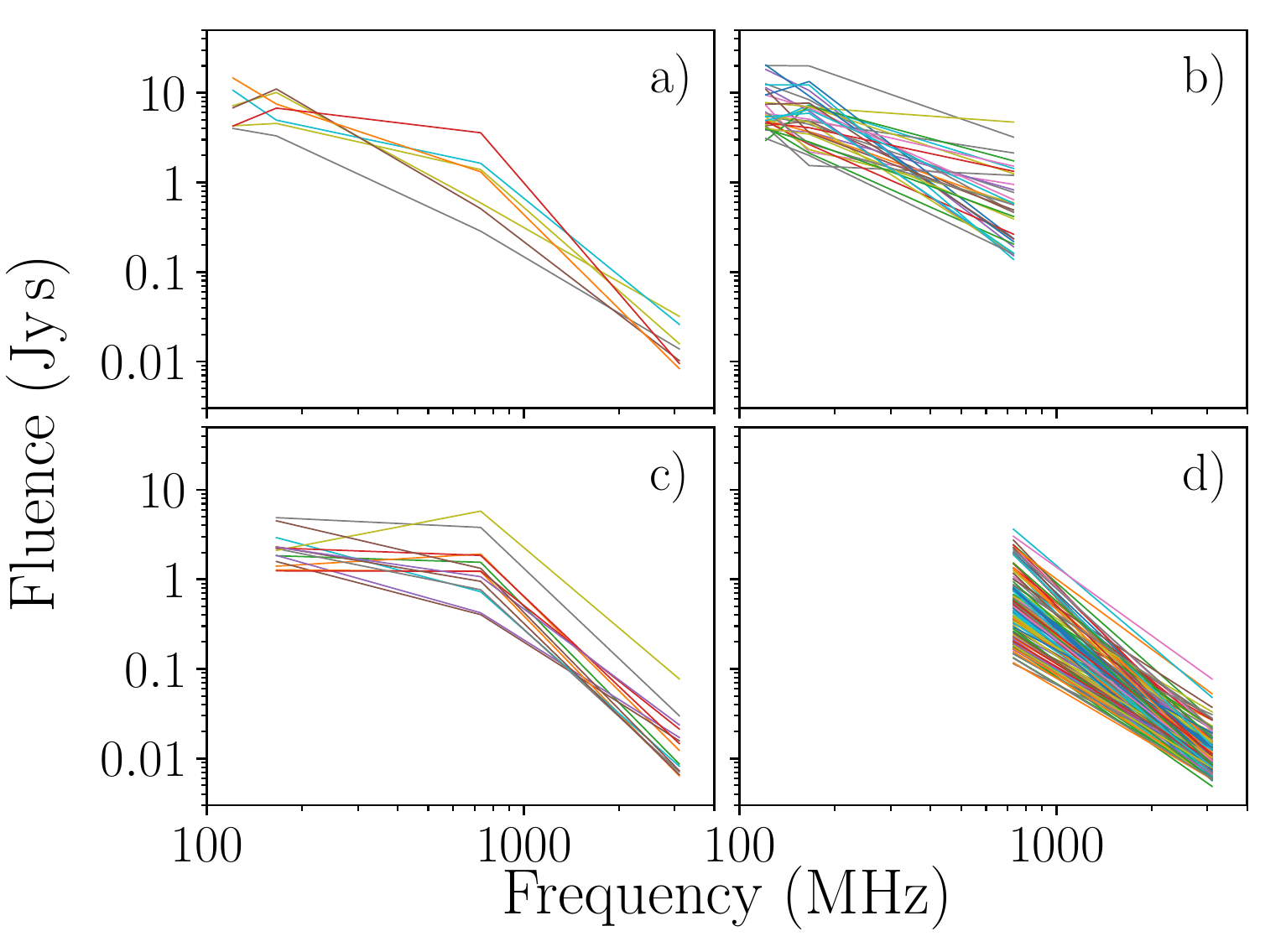}
\caption{Selected samples of giant pulses based on their simultaneous detections.
Giant pulses with simultaneous detections in all four bands are plotted in a).
Giant pulses detected simultaneously without a 3100\,MHz detection are shown in b), while those pulses with only a 3100, 732 and 165.76\,MHz simultaneous detection are shown in c).
Panel d) contains giant pulses only detected between 732 and 3100\,MHz.}
\label{fig:alpha_examples}
\end{figure}

\begin{figure}
\includegraphics[width=\linewidth]{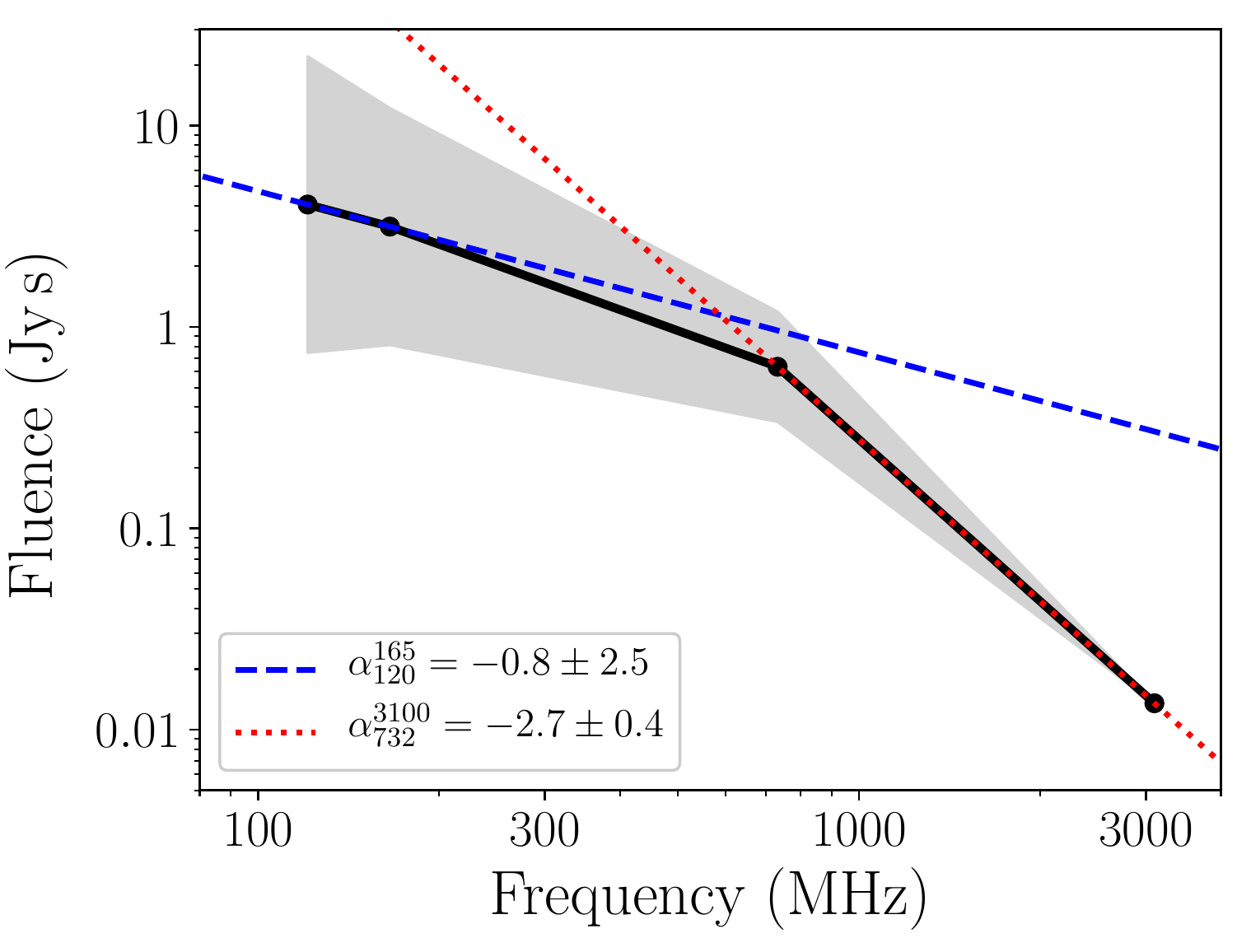}
\caption{An archetypal average spectrum of the detected giant pulses.
Each spectral point is calculated based on the mean spectral index between the two frequencies. 
This is the expected shape of a giant pulse spectrum, for a reference value of 0.013\,Jy\,s at 3100\,MHz.
The gray shaded error region represents the median absolute deviation for each spectral index distribution.
The power-law fits are based on only the two Parkes bands (red dotted) and on the two lowest MWA bands (blue dashed).}
\label{fig:synth_spec}
\end{figure}

In Figure~\ref{fig:alpha_examples} we plot three different samples of giant pulses, based on the frequency bands in which they were detected.
In general, these spectra also show a tendency of flattening at the lower frequencies.
An archetypal synthetic giant pulse spectrum based on the spectral index distributions is shown in Figure~\ref{fig:synth_spec}, which demonstrates the expected pulse spectral shape given a 3100\,MHz fluence of 0.013\,Jy\,s.
The shaded error region is calculated using the median absolute deviation of the spectral index distribution, instead of the standard deviation, as it is less sensitive to the existence of extreme values (see Figure~\ref{fig:alpha_dists}).
The power-laws drawn are fits to the two Parkes bands and the 165.76 and 120.96\,MHz MWA subbands.

The mean spectral index between 3100 and 732\,MHz from the synthetic spectrum is $ \alpha_{732}^{3100}=-2.7 $ with a width of 0.4.
Between 732 and 165.76\,MHz the synthetic spectral index becomes shallower with $ \alpha_{165}^{732}=-1.1 $ and a width of 0.4.
Between 165.76 and 120.96\,MHz is estimated to be $ \alpha_{120}^{165}=-0.8 $ with a distribution width of 2.5.
The large error in $ \alpha_{120}^{165} $ is due to a combination of relatively large errors in fluence estimates and that the frequencies are relatively close together, hence there is a wide distribution of spectral indices and therefore a less well constrained mean. 
Spectral index information between each of the bands and from the synthetic spectrum are shown in Table~\ref{tab:spec_index}.

\begin{figure}
\includegraphics[width=\linewidth]{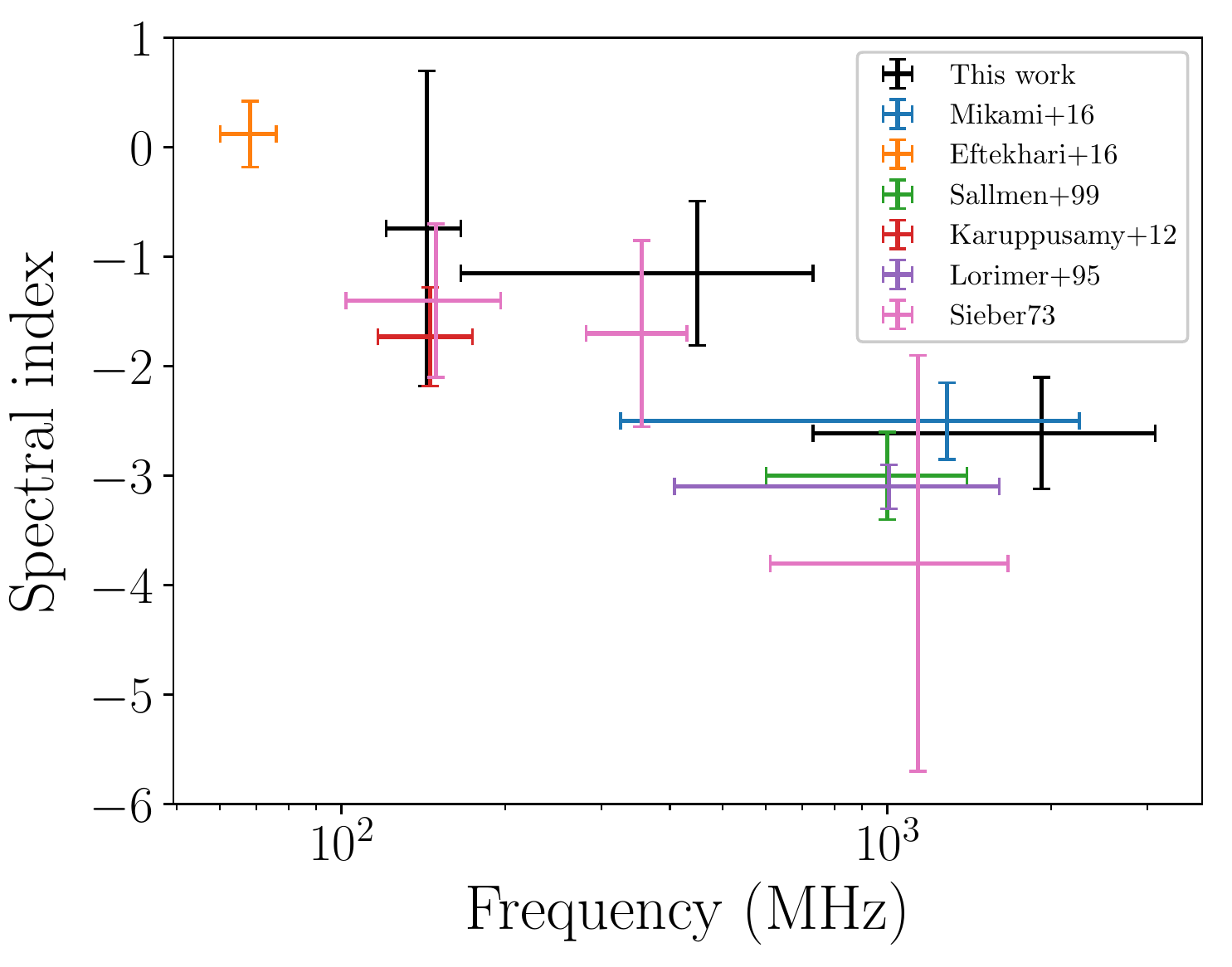}
\caption{A comparison of our measured spectral indices with those reported in the literature. 
Our data (black) are from simultaneous observations and shows the spectral behavior over a wide frequency range.
The spectral indices from the literature, by and large, are consistent with the spectral flattening that is indicated by our observations.
The horizontal bars represent the frequency range over which the spectral indices were calculated. 
The vertical bars represent the errors reported by the sources, which in most cases corresponds to the range of indices possible based on flux density errors.
The distribution widths, or $ \pm 50\% $, were used in the cases where no error/range information was available.}
\label{fig:spectral_indices}
\end{figure}

The synthetic spectrum, in addition to the spectral index histograms and fluence distribution clustering, is evidence for a flattening spectrum for giant pulses at low frequencies.
Moreover, we gathered measurements of spectral indices from the literature \citep{1973AASSieber,1995MNRASLorimer,1999ApJSallmen,2012AAKaruppusamy,2016ApJEftekhari,2016ApJMikami} and compared them to our measurements (see Figure~\ref{fig:spectral_indices}). 
We find that our results are consistent with previous measurements of the giant pulse spectral index.

\begin{table*}[htbp]
\renewcommand{\arraystretch}{1.2}
\centering
\caption{Spectral index distribution and synthetic spectrum parameters. \label{tab:spec_index}}
\begin{tabular}{ccc}
\hline
& Measured\tablenotemark{$a$} & Synthetic\tablenotemark{$b$} \\
\hline\hline
$ \alpha_{120}^{165} $ & $ -0.74^{+1.4}_{-1.8} $ & $ -0.79\pm2.5 $\\
$ \alpha_{165}^{732} $ & $ -1.15^{+0.8}_{-0.6} $ & $ -1.07\pm0.4 $\\
$ \alpha_{732}^{3100} $ & $ -2.61^{+0.5}_{-0.7} $ & $ -2.66\pm0.4 $\\
\hline
$ \alpha_{185}^{732} $ & $ -0.78^{+0.73}_{-0.72} $ & --\\
\hline
\end{tabular}
\tablenotetext{a}{The quoted uncertainties represent the 12.5 and 87.5 percentile (i.e. where 75\% of pulses are present about the mean).}
\tablenotetext{b}{The errors represent the distribution width only.}
\end{table*}

The average main pulse spectral index we find between the two Parkes bands (Figure~\ref{fig:alpha_dists}) is consistent with the value ($ \alpha=-2.4\pm0.5 $) computed by \citet{2016ApJMikami}, however the interpulse spectral index is difficult to compare given the small number of pulses detected simultaneously.
The spectral index we calculate between the Parkes 732 and MWA 165.76\,MHz subbands is consistent with the shallower value ($ \alpha=-1.7\pm0.5 $) calculated by \citet{2010AAKaruppusamy}.

As a test, we supposed our fluence estimates were significantly in error and that the spectral index is in reality $ \alpha=-2.7 $, even over our low frequency subbands.
With a reference fluence of 0.013\,Jy\,s at 3100\,MHz, as in Figure~\ref{fig:synth_spec}, this would require a mean fluence of $ \sim 83 $\,Jy\,s at 120.96\,MHz, $ \sim 35 $\,Jy\,s at 165.76\,MHz, and $ \sim 19 $\,Jy\,s at 210.56\,MHz, overestimating the average fluences by a factor of $ \sim 10 $ based on the values recorded in Table~\ref{tab:GPfluences}.
Additionally, if we assume the distributions follow the same power-law behavior and that our noise statistics would remain unchanged, then we can calculate the number of detectable giant pulses ($ N_1 $) above the extrapolated median fluences ($ F_1 $) using
\begin{equation}\label{eq:pl_test}
N_1 = N_0\left(\frac{F_0}{F_1}\right)^{-\hat{\beta}},
\end{equation}
where $ N_0 $ is the measured number of pulses above the measured median fluence $ F_0 $, and $ \hat{\beta} $ is the measured power-law exponent.
Computing this for each of the three MWA subbands, for main pulses only, yields an expected $ \sim 5\times 10^6 $ detectable pulses at 210.56\,MHz, $ \sim 4\times 10^5 $ at 165.76\,MHz, and $ \sim 9\times 10^5 $ at 120.96\,MHz.
These predictions are between a factor of $ \sim 10^{3\text{--}4}$ times larger than the recorded numbers of main pulses.
It is therefore implausible that the spectrum continues with the steep index to low frequencies.

\subsection{Non-giant pulse emission}
For the Parkes data, we also attempted to recover the non-giant pulse emission from the Crab.
For this, we essentially treated all pulses with a detection below a $ 3.5\sigma $ threshold as being ``non-giant pulse'' emission.
All such pulses were synchronously averaged to construct an ``integrated profile''.
At 732\,MHz, the MP and IP components of such a profile are approximately equal in amplitude ($ S_\mathrm{peak}\sim 19 $\,Jy), whereas in the constructed giant pulse profile (detections $ \geq 6\sigma $), the MP is $\sim 6$ times brighter than the IP.
At 3100\,MHz, the giant pulse profile is dominated by the MP emission, and there is only a marginal peak at the IP phase. 
The non-giant pulse profile at this frequency contains both MP and IP components, though the MP ($ S_\mathrm{peak}\sim 150 $\,mJy) is only $ \sim 2 $ times brighter than the IP. 
Based on this, if we calculate the spectral index for the MP ($ \alpha_\mathrm{MP} $) and IP ($ \alpha_\mathrm{IP} $) non-giant pulse emission, we find that $\alpha_\mathrm{MP}\approx -3.3\pm0.1$ and $\alpha_\mathrm{IP}\approx -3.8\pm0.1$.
In comparison with the published estimates of the normal emission spectral behavior (e.g. \citealp{1999ApJMoffett}; $\alpha_\mathrm{MP}=-3.0$ and $\alpha_\mathrm{IP}=-4.1$), we find that our results appear consistent.

We, however, did not carry out such an analysis on the MWA subbands because of the severity of the pulse broadening (see Section~\ref{sec:scattering}), which makes it extremely difficult to disentangle the non-giant pulse emission from weak, scattered giant pulses.

\section{DISCUSSION}\label{sec:discussion}

\subsection{Spectral flattening}\label{sec:gp_spec_flatten}
Our analysis identifies a spectral flattening at low frequencies in Crab giant pulses.
A flattening spectrum was also hinted at by \citet{2015ApJOronsaye}, whose analysis showed the spectrum becomes shallower by $ \sim 5\% $ at lower frequencies based on Monte Carlo simulations of observations at 193\,MHz and 1382\,MHz.
We note, however, that the fluences presented by \citet{2015ApJOronsaye} are significantly different (by orders of magnitude) to those we calculate here.
Re-examining the Parkes data used, we estimate that the flux densities are a factor of $ \sim 10\text{--}100 $ larger than quoted and attribute this to an error in the flux density calibration in the original processing.
This discrepancy is also noted by \citet{2016ApJMikami}, whose observing bands are at a similar frequency to those used by \citet{2015ApJOronsaye}.
The MWA fluences we calculate herein are roughly consistent with the estimates made by \citet{2015ApJOronsaye}, which together with the re-evaluated Parkes fluences implies that the flattening observed is more significant than the authors stated. 

The two power-law slopes we identify behave similarly to those broken-type spectra \citep{2000AASMaron,2013MNRASBates}, where $ |\alpha_\mathrm{low}|<|
\alpha_\mathrm{high}| $.
The average spectral indices we see from our giant pulse sample ($ \alpha_{120}^{165}=\alphamwalomi $, $ \alpha_{165}^{732}=\alphamipkslo $ and $ \alpha_{732}^{3100}=\alphapkslohi $) are consistent with the estimates of \citet{2000AASMaron} for normal pulsar emission, $ \langle\alpha_\mathrm{low}\rangle=-0.9\pm 0.5 $ and $ \langle \alpha_\mathrm{high}\rangle=-2.2\pm 0.9 $.
\citet{2016ApJMikami} report a main pulse spectral index between 325 and 2250\,MHz of $ \alpha_{325}^{2250}=-2.44\pm0.47 $, which is consistent with our estimated main pulse high-frequency spectral index.

We acknowledge that given we have only 4 spectral points, thus there is not enough information to robustly determine the actual spectral index values and associated uncertainties in the synthetic spectrum.
The uncertainty in the MWA fluences is generally the most significant source of error, especially at the lowest frequency where the pulses tend to be scattered, and appropriately characterizing the pulses is difficult.

There is an increasing amount of evidence for a slightly flatter, or even an inverted spectrum at low frequencies (e.g. \citealp{2007ApJBhat,2010AAKaruppusamy,2015ApJOronsaye,2016ApJEftekhari}).
In contrast, \citet{2006ARepPopov} calculate giant pulse spectral indices between $ -3.1 $ and $ -1.6 $ for 111--600\,MHz and $ -3.1 $ to $ -2.5 $ for 23--111\,MHz, both with a mean of $ -2.7\pm0.1 $, however note that these values are subject to selection effects.
In addition to this, their errors in fluence and spectral index are likely optimistic given that at 23\,MHz the giant pulse rise time alone would be several tens or hundreds of pulse periods. 

While there is indeed a wide spread in the spectral indices quoted in the literature, the general trend is a shallower spectral index at low frequencies. (see Figure~\ref{fig:spectral_indices}).
Since our data are from simultaneous observations, we are able to confidently assert that the spectral index tends to be shallower at low frequencies.
If we only use the values from the literature, a direct comparison is difficult as they are from different instruments and measured at widely separated epochs (sometimes spanning decades).

The implications for the giant pulse emission mechanism is that we would need some process or propagation effect (possibly within the magnetosphere) that allows for a flattening and eventual turn-over (which likely occurs at $ \nu \ll 100 $\,MHz) in the spectrum.
As with the GPS pulsars, this effect is perhaps caused by the surrounding environment of the pulsar (i.e. the Crab nebula in this case).
However, \citet{2015ApJOronsaye} showed that at MWA frequencies, free-free absorption from within the nebula (e.g. \citealp{1997ApJBietenholz}) is not able to explain the flattening they observe, with free-free absorption coefficients on the order of $\sim10^{-23}\,\mathrm{cm^{-1}}$.
Given our flattening is more apparent than represented previously, free-free absorption alone causing the flattening is unlikely.
Structures in the nebula and the intervening ISM (e.g. \citealp{2011ApJSmith}) may be capable of attenuating the fluence estimates by a few percent, but would require 10--100 such filaments to be intercepted.
Not only is the chance alignment of filaments unlikely, but the DM of the pulsar would be increased by $ \sim\mathrm{few\,pc\,cm^{-3}} $ which is unphysical.

\subsection{Emission mechanism}\label{sec:gp_emission}
The giant pulse fluence dependence on frequency, particularly the flattening at low frequencies, is not predicted in detail in any of the current models. 
The spectral behavior provides important information about what physical processes are producing the emission.

The coherent radio emission mechanism for pulsars is still unknown (see e.g. \citealp{1995JApAMelrose} for a review), especially given the complexity of modeling pulsar magnetospheres (e.g. \citealp{2006ApJSpitkovsky,2012ApJLi,2013MNRASTchekhovskoy}) and the myriad emission models in the literature.
There are several models that are able to address individual aspects of giant pulse emission \citep{2016JPlPhEilek}, though none are able to explain all of the characteristics alone, possibly because they are not fully explored in the non-linear regime (see e.g. \citealp{2002nsps.confEilek}).

Main pulse emission from the Crab is comprised of narrow-band nanoshots (e.g. \citealp{2007ApJHankins,2016ApJHankins}).
The emission we observe is the average of many of these nanoshots, where the center frequency depends on which emission model is selected.
We examine two plasma emission models, following \citet{2016JPlPhEilek}.

Strong plasma turbulence \citep{1997ApJWeatherall,1998ApJWeatherall} relies on relativistic particles driving the production of plasma waves which are converted into electromagnetic radiation and escape to potentially produce nanoshots.
The emission is produced at a frequency $ \nu_\mathrm{SPT} \sim 2\gamma_\mathrm{s}^{1/2}\nu_\mathrm{p} $, where $ \gamma_\mathrm{s} $ is the Lorentz factor describing the streaming speed of the pair plasma and $ \nu_\mathrm{p} $ is the plasma frequency.
In order for this emission be observed in the radio, the plasma densities must be enhanced by a factor of $ 10^2\lesssim \lambda\gamma_\mathrm{s}\lesssim 10^5$, where $ \lambda = n/n_\mathrm{GJ} $ and $ n_\mathrm{GJ} \sim 10^6\text{--}10^7\mathrm{\,cm^{-3}} $ is the Goldreich-Julian (GJ) density.
The flux densities of the nanoshots are predicted to scale with frequency as $ S_\nu\propto \nu^{-1} $, assuming radius-to-frequency mapping and ignoring effects related to polar cap current flow.

A free-electron maser model involves the interaction of relativistic particle beams with plasma waves to induce charge bunching, leading to strong coherent bursts of radiation.
The emission frequency, assuming the plasma is at rest (e.g. \citealp{1992ITPSBenford}), is $ \nu_\mathrm{FEM}\sim 2\gamma_\mathrm{b}^2\nu_\mathrm{p} $, where $ \gamma_\mathrm{b} $ describes the speed of the driving particle beam.
For radio frequency emission, this requires a density enhancement similar to that of the strong plasma turbulence, $ 10^2\lesssim \lambda\gamma_\mathrm{b}^4\lesssim 10^5$. 

The flattening spectrum then raises the question of what is driving the nanoshot emission in the regions where conditions translate to emission at low frequencies.
Crab giant pulse radio emission is suspected to originate higher in the magnetosphere, perhaps near the light-cylinder.
This is based on the relative enhancements required for radio emission in comparison to pair-production plasma models (e.g. \citealp{2002ApJArendt,2016JPlPhEilek}).
High-altitude emission is also supported by multi-wavelength observations of the Crab identifying that the high-energy and radio profiles are very close in pulse longitude, implying they originate from similar regions within the magnetosphere \citep{2010ApJAbdo}.
While the strong plasma turbulence model has a shallow predicted scaling for nanoshot flux density which supports a flatter spectrum, it is unclear how that scaling translates into the regime where we are observing the superposition of many nanoshots.
If $ S_\nu\propto\nu^{-1} $ is representative for unresolved emission, then the model is unable to explain the steep spectral index typically observed above $ \sim 300 $\,MHz, even though the model is able to describe the nanoshot time scales and frequency structure.

If we assume that in fact both phenomena are present within the magnetosphere, then the relative dominance of the processes would depend on, for example, the driving beam densities and ambient plasma characteristics.
Typically, one can assume that the charged particles streaming from the pulsar are accelerated along the electric fields as they move away from the neutron star surface.
In most models, $ \gamma_\mathrm{s}>100 $ and $ \gamma_\mathrm{b}^2\sim 10\text{--}100 $ are required in order to match the observed nanoshot frequency-time product \citep{2016JPlPhEilek}.
In this way, one could imagine strong plasma turbulence begins to dominate in the region where the low radio frequency emission is produced in the upper magnetosphere, where particles are further away from the star and therefore traveling faster.

Without further exploration of these models (and others), in terms of observational emission characteristics, it is difficult to say more.
How the nanoscale attributes translate to millisecond time scales, and predictions for the flux density frequency scaling, are critical for meaningful comparison to observation.
At low frequencies, there is the additional complication of pulse broadening which distorts the intrinsic emission.

\subsection{FRBs as extragalactic super-giant pulses}\label{sec:gp_frbs}
Wide band observations are able to provide limits of FRB spectral index distributions (e.g. \citealp{2016ApJBurke-Spolaor}). 
Typically, the measured spectral indices of FRBs are poorly constrained.
For example, the measured spectral index (1.214--1.537\,GHz) of FRB 121102 ranges between $-10$ and $+14$ \citep{2016NatSpitler}, and for other FRBs the range is approximately $-8$ to $+6$ (e.g. \citealp{2007SciLorimer,2012MNRASKeane,2015ApJRavi,2016ApJBurke-Spolaor}).
These values are consistent with the large spread in spectral indices measured for Crab giant pulses, including those calculated herein.
If some FRBs are ``super-giant'' pulses from extragalactic pulsars, and assuming our low-frequency spectral index ($ \alpha=\alphamwalomi $) is representative, it is possible to estimate the number of expected FRB detections at MWA frequencies.
Based on the calculations of \citet{2013ApJTrott}, we would expect to see somewhere between $ \sim 0.1\text{--}100 $ FRBs per 10-hours of observing with the MWA, above a signal-to-noise ratio of 7, depending on scattering effects and the data processing.

Given that no low-frequency instrument has claimed an FRB detection to date (e.g. \citealp{2014AACoenen,2015AJTingay,2015MNRASKarastergiou,2016MNRASRowlinson}), there are two obvious constraints we can make.
If FRBs are close enough to be detectable ($\lesssim \mathrm{few\ hundred\ Mpc} $), then the non-detections thus far would suggest that the spectrum has turned over or flattened sufficiently for the giant pulses to become undetectable.
From our results, this seems at least plausible assuming that the emission originates from a Crab-like pulsar.
However, if the spectrum has not inverted then the non-detections perhaps suggest that these objects are much further away than assumed in the giant pulse FRB models.
The latter is supported by the localization of FRB 121102 \citep{2017NatChatterjee} at $\sim$1\,Gpc and the stable DM that FRB 121102 exhibits (see e.g. \citealp{2017ApJLLyutikov}).
With these results in mind, a ``super-giant'' pulse origin for FRBs seems less likely.

\section{CONCLUSIONS}\label{sec:conclusion}
We have reported on simultaneous observations of the Crab pulsar conducted with the MWA (120.96, 165.76, 184.96 and 210.56\,MHz) and Parkes radio telescope (732 and 3100\,MHz).
Our observations sampled from 120 to 3100\,MHz (a factor of $ \sim 30 $ in frequency), and thus simultaneously span low-, mid- and high-frequencies, which provides a unique view of the giant pulse spectrum.
Giant pulses were detected in all bands, ranging from \nMWAlo\ at 120\,MHz to \nPKSlo\ at 732\,MHz. 
Seven giant pulses (6 main pulses and 1 interpulse) were detected simultaneously in five of the observing bands (excluding 184.96\,MHz due to no time overlap with the 120.96, 165.76, and 210.56\,MHz bands).
The correlation of detected pulses between bands varies, ranging from $ \sim 40\% $ (184.96 to 732\,MHz, relative to 184.96\,MHz detections) to $ \sim 87\% $ (120.96 to 165.76\,MHz, relative to 120.96\,MHz detections).

The mean spectral index for the sample of simultaneous giant pulses tends to flatten at low frequencies, from $ \alpha=\alphapkslohi $ (732--3100\,MHz) to $ \alphamwalomi $ (120.96--165.76\,MHz). 
By creating a synthetic spectrum based on the distributions of spectral indices, we also see the evolution in spectral shape is not well characterized by a single power-law.
Furthermore, we compare our simultaneous wideband results with spectral index measurements from the literature, which further reinforces the observed spectral flattening.
This flattening is unlikely to be caused by propagation effects within the nebula.

The emission mechanism required to explain this phenomenon is currently not well understood.
Further work is required to extend current giant pulse emission models in order to determine how the flux density spectrum changes and how the intrinsic nanoshot characteristics translate to observing their superposition.

We also measured the characteristic pulse broadening times for giant pulses in the MWA subbands.
Specifically, we calculated a frequency scaling index of $ \alpha_\mathrm{d}=-3.7\pm0.4 $ which is consistent with the literature relating to the scattering characteristics of Crab giant pulses.

We also comment on the plausibility of a giant pulse origin of some FRBs.
Considering the localization of FRB 121102, and the flattening spectrum that we observe, it appears that a giant pulse emission origin for FRBs (assuming the Crab is typical) is less likely.
This is supported by the non-detections of FRBs from any low frequency telescope to date.

Investigations of giant pulse spectra over wide frequency ranges, especially extending down below $ \sim 100 $\,MHz, have not been attempted for other giant-pulse emitting pulsars.
Such studies are particularly important to check whether the Crab is a special case or typical in terms of giant pulse emission.
We also emphasize the important role of simultaneous observations in this endeavor. 

\acknowledgments
The authors would like to thank the anonymous referee for their valuable input and suggestions, which have significantly improved the paper. 
B. W. M. would like to thank A. Sutinjo, D. Ung and M. Xue for useful discussions regarding array factor calculations, array efficiencies and flux density calibration.
N. D. R. B. acknowledges the support from a Curtin Research Fellowship (CRF12228).
F. K. acknowledges funding through the Australian Research Council grant DP140104114.
The authors acknowledge the contribution of an Australian Government Research Training Program Scholarship in supporting this research. 

This scientific work makes use of the Murchison Radio-astronomy Observatory, operated by CSIRO. 
We acknowledge the Wajarri Yamatji people as the traditional owners of the Observatory site. 
Support for the operation of the MWA is provided by the Australian Government (NCRIS), under a contract to Curtin University administered by Astronomy Australia Limited. 
We acknowledge the Pawsey Supercomputing Centre which is supported by the Western Australian and Australian Governments. 
The Parkes Radio Telescope is part of the Australia Telescope National Facility, which is funded by the Commonwealth of Australia for operation as a National Facility managed by CSIRO.
Parts of this research were conducted by the Australian Research Council Centre of Excellence for All-sky Astrophysics (CAASTRO), through project number CE110001020. 
Some of the results in this paper have been derived using the HEALPix (K. M. G\'{o}rski et al., 2005, ApJ, 622, p759) package.

\bibliography{references}
\bibliographystyle{aasjournal.bst}

\appendix
\section{Array factor calculation}\label{appendix:arrayfactor}
An antenna element in isolation has a complex voltage pattern given by some frequency-dependent function $ D(\theta,\phi) $, where $ \theta $ is the zenith angle and $ \phi $ is the azimuth.
The function $ D(\theta,\phi) $ is called the \textit{element pattern} and gives the signal strength received by the element for any given direction, assuming it is positioned at the origin, $ \textbf{r}=(0,0,0) $.
The coordinate system used here is such that the azimuth ($ \phi $) is defined with $ 0^\circ $ directly East and increases in an anticlockwise direction. 
The zenith angle ($ \theta $) is defined in the normal convention.

For an array of $ N $ elements, we define each element voltage pattern as $ D_n(\theta,\phi) $.
The tied-array beam pattern will be the sum of each element pattern in response to a wave, $ \psi_n $, impinging on the array.
Given that the source is in the far-field, this wave will be planar.
It is practical to express the planar wave in terms of the coordinate system we have adopted, thus $ \psi_n $ can be written as
\begin{equation}
\psi_n = \exp\left(i\textbf{k}\cdot\textbf{r}_n\right) \equiv \exp\left[\frac{2\pi i}{\lambda}\left(x_n\sin\theta\cos\phi + y_n\sin\theta\sin\phi + z_n\cos\theta\right)\right],
\label{appeq:planewave}
\end{equation}
where $ \textbf{k} $ is the three-dimensional wave vector, $ \textbf{r}_n=(x_n,y_n,z_n) $ is the position of the $n$th element relative to the center of the array and $ \lambda $ is the observing wavelength. 
 
We also apply weights, $ w_n $, on a per element basis. 
For the MWA, when calculating the beam pattern for an individual tile (which consists of 16 dipole elements), these weights incorporate information about the cable losses and port currents required to accurately model the mutual coupling between dipoles and polarization characteristics \citep{2015RaScSutinjo}.
On the tied-array scale, each element is now one MWA tile and the weights encode the phase delay information required to correctly point the array at a given sky position. 

The tied-array voltage pattern is
\begin{equation}
D_\mathrm{array}(\theta,\phi) = \frac{1}{N}\sum\limits_{n=1}^{N} w_n D_n(\theta,\phi) \psi_n.
\label{appeq:array_pattern}
\end{equation}
If we assume the array elements are identical, then one can move the element factor out of the summation, and equation~(\ref{appeq:array_pattern}) becomes
\begin{equation}
D_\mathrm{array}(\theta,\phi) = D(\theta,\phi)\frac{1}{N}\sum\limits_{n=1}^{N} w_n \psi_n.
\label{appeq:array_pattern_ident}
\end{equation}
Given we have two separable factors in equation~(\ref{appeq:array_pattern_ident}), one of which is the element pattern, we define the other as the \textit{array factor},
\begin{equation}
f(\theta,\phi) = \frac{1}{N}\sum\limits_{n=1}^{N} w_n \psi_n(\theta,\phi).
\label{appeq:array_factor}
\end{equation}
The array factor represents the response of an array of identical elements and encompasses the interference effects from the individual element patterns in response to the received radiation from the visible sky.

To point the tied-array radiation pattern, we adjust the complex weights $ w_n $. 
In this case, we require the array factor to be unity at the desired pointing center, thus the weights are expressed as the complex conjugate of $ \psi_n $ evaluated only at the target position.
Thus, the array factor pointed at some target zenith angle ($ \mathrm{za} $) and azimuth ($ \mathrm{az} $), is given by
\begin{equation}
f(\theta,\phi;\mathrm{za},\mathrm{az}) = \frac{1}{N}\sum\limits_{n=1}^{N} \psi_n(\mathrm{za},\mathrm{az})^\dagger \psi_n(\theta,\phi)
\label{appeq:pointed_array_factor}
\end{equation}
where $ \psi_n^\dagger $ denotes the complex conjugate of $ \psi_n $.
This ensures that the array factor power pattern, $ |f(\theta,\phi)|^2 $, will be unity only at the pointing center, and in the range $ [0,1) $ elsewhere.
The phased array power pattern is then
\begin{equation}
B_\mathrm{array}(\theta,\phi)=|D_\mathrm{array}(\theta,\phi)|^2 = |D(\theta,\phi)|^2|f(\theta,\phi)|^2,
\label{appeq:tied-array_beam}
\end{equation}
which is evaluated over $ \theta=[0,\pi/2] $ and $ \phi=[0,2\pi) $ to recover the array response to the sky visible to the elements.
Both the element factor and array factor are also functions of frequency, $ \nu $, therefore the tied-array beam pattern is a function of frequency and direction.

This process effectively recreates the naturally weighted synthesized beam for the array.
The element pattern, $ D(\theta,\phi) $, for the MWA has a grid-like morphology due to the MWA tiles being a regularly spaced grid of dipoles, thus we find that for some frequency and pointing combinations the tile pattern side lobes can have similar, or exceed the sensitivity of the main lobe.
For a pseudo-random array, the tied-array beam pattern grating lobes will be randomly distributed across the sky for each pointing and frequency, thus the element pattern dominates the sensitivity pattern on the sky.
Contrary to our assumption, each tile is not necessarily identical, with some instances of individual dipoles failing which reduces the tile sensitivity by $ \sim 1/16 $.
This effect is not accounted for in the beam simulations.

\begin{figure*}
\centering
\includegraphics[width=0.45\linewidth]{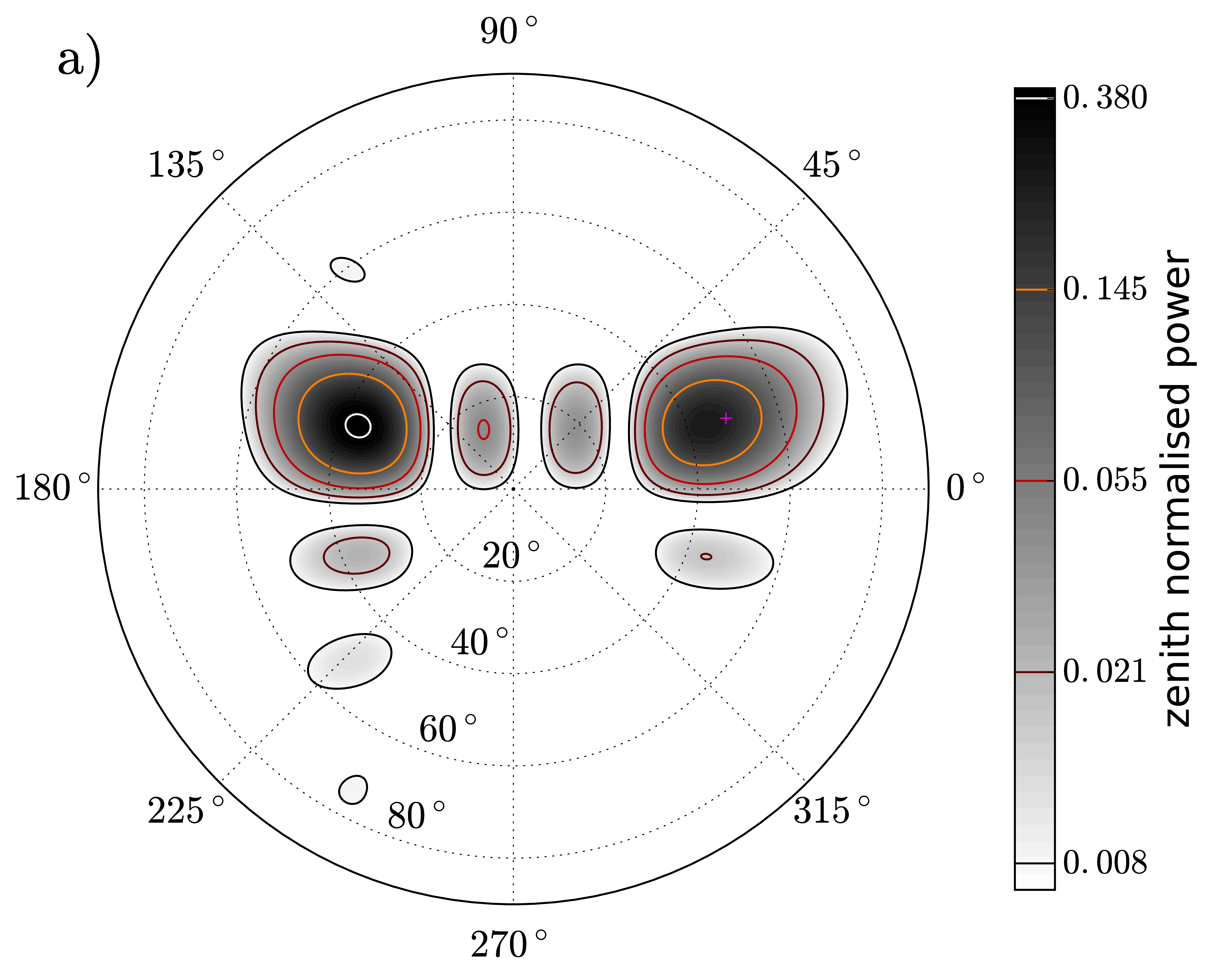} 
\includegraphics[width=0.45\linewidth]{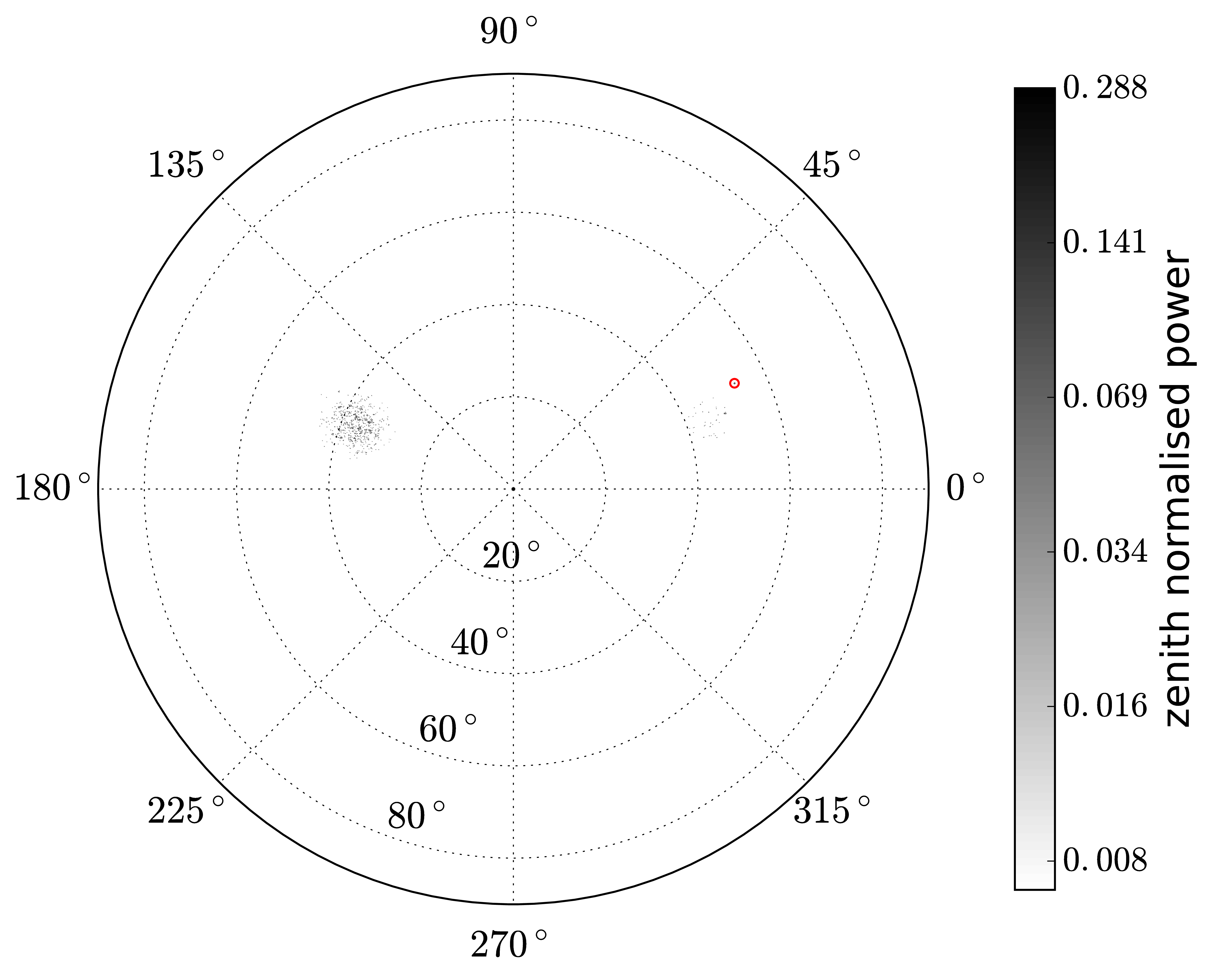} 
\rule{\textwidth}{1pt}
\includegraphics[width=0.45\linewidth]{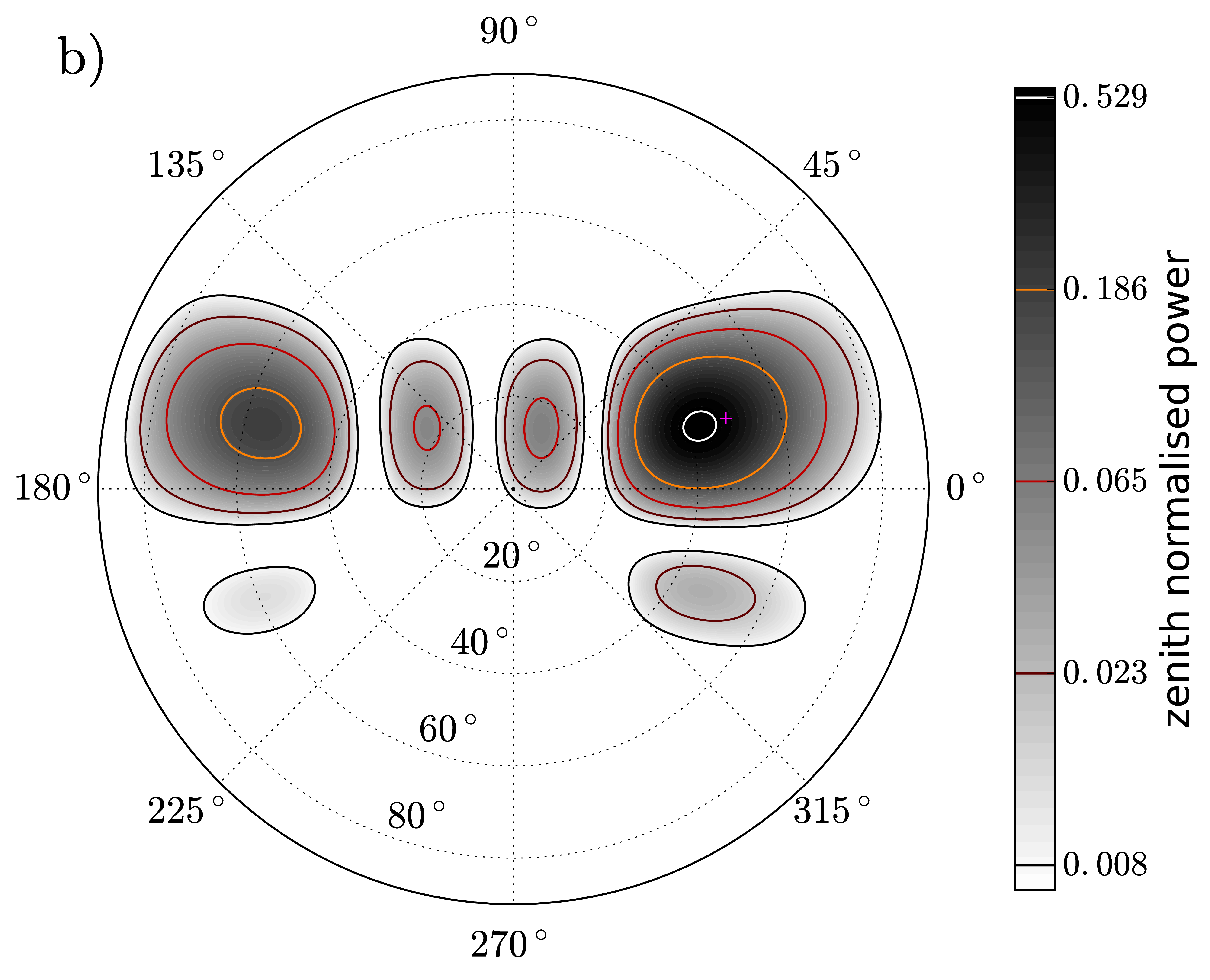} 
\includegraphics[width=0.45\linewidth]{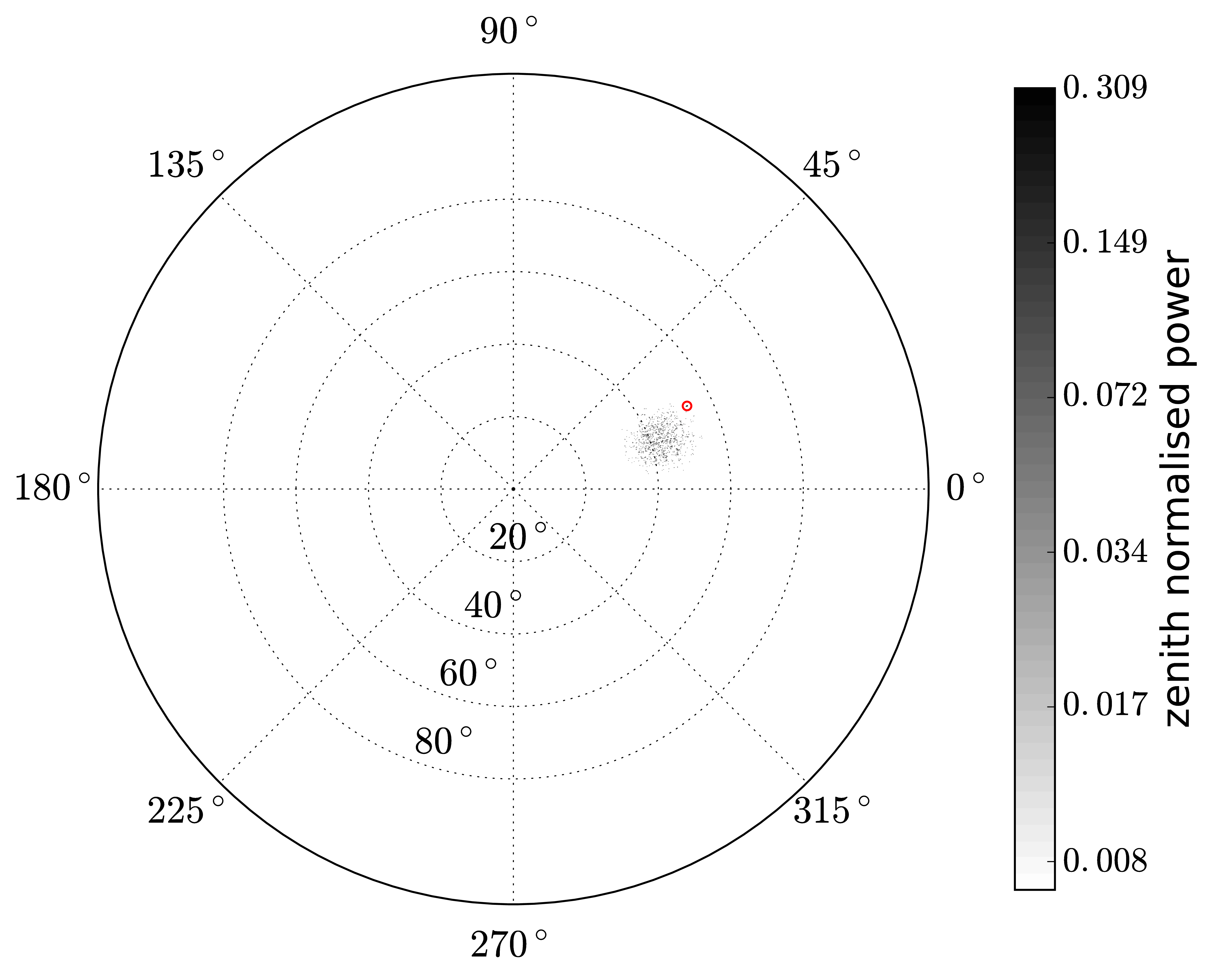} 
\rule{\textwidth}{1pt}
\includegraphics[width=0.45\linewidth]{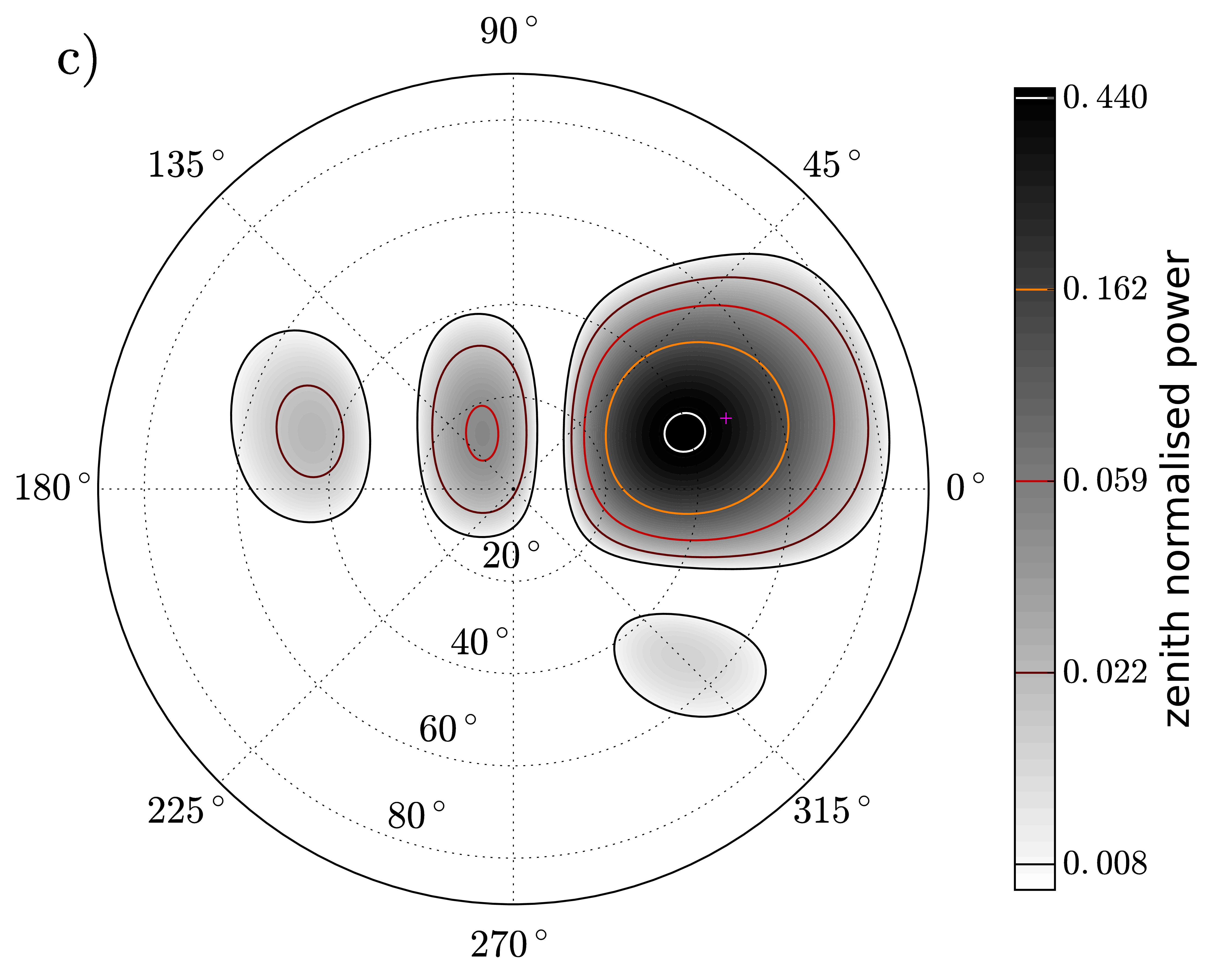} 
\includegraphics[width=0.45\linewidth]{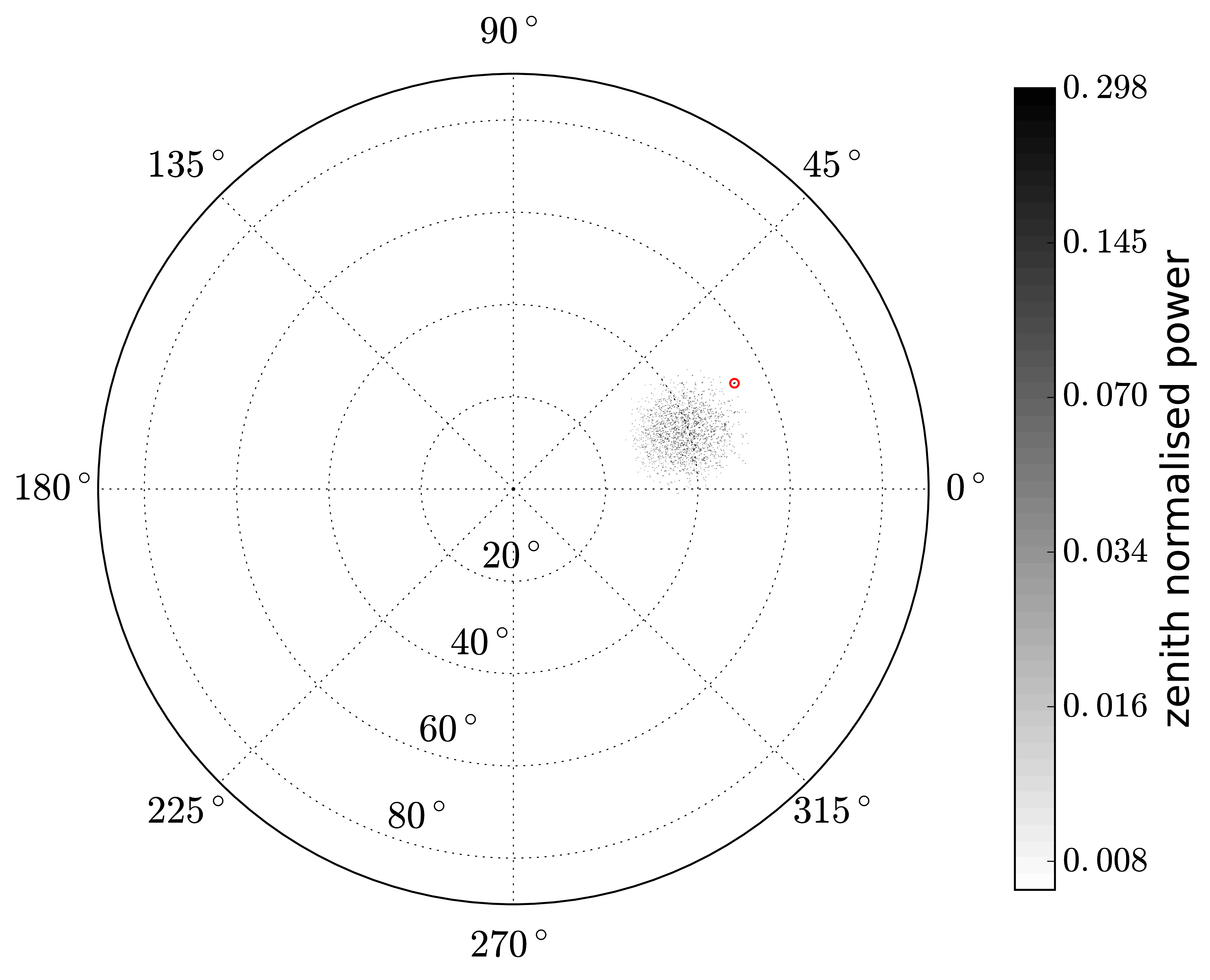} 
\caption{The MWA tile pattern (left) and tied-array beam pattern (right) for each frequency: a) $ 210.56\mathrm{\,MHz} $, b) $ 165.56\mathrm{\,MHz} $, and c) $ 120.96\mathrm{\,MHz} $.
The gray-scale background gradient and the colored contours denote the zenith normalized power for the beam.
The magenta cross marks the tile beam pointing center ($ \mathrm{azimuth}=18.43^\circ $, $ \mathrm{zenith\ angle}=48.57^\circ $).
In the case of the $ 210.56\mathrm{\,MHz} $ beam, the highest tile beam sensitivity region actually exists in the side-lobe.
The red circles highlight the target position on each of the tied-array beam patterns.
\label{fig:tile_beams}}
\end{figure*}

As an example, Figure~\ref{fig:tile_beams} shows a simulated MWA tile beam pattern and tied-array beam pattern at 210.56, 165.76 and 120.96\,MHz.
An important note here is that both the tile beam and tied-array beam models are theoretical, and in reality the true beam patterns will have features not described here.

\subsection{Antenna temperature}\label{appendix:1:tsys_calc}
The antenna temperature $ T_\mathrm{ant}(\nu,\theta,\phi) $ is calculated as the product of the antenna pattern $ B_\mathrm{array}(\nu,\theta,\phi) $ and the sky temperature $ T_\mathrm{sky}(\nu,\theta,\phi) $ via the convolution
\begin{equation}
T_\mathrm{ant}(\nu,\theta,\phi)=\dfrac{\displaystyle\int_{4\pi}B_\mathrm{array}(\nu,\theta,\phi)T_\mathrm{sky}(\nu,\theta,\phi) \,\mathrm{d}\Omega}{\displaystyle\int_{4\pi} B_\mathrm{array}(\nu,\theta,\phi)\,\mathrm{d}\Omega}.
\label{appeq:tant}
\end{equation}
The tied-array beam pattern was output in the necessary format for software used by \citet{2015PASASokolowski} to compute the above integral with the GSM, which is natively produced in HEALPix\footnote{http://healpix.sourceforge.net/} format.

\subsection{Tied-array gain}\label{appendix:2:gain_calc}
To calculate the tied-array gain, we first determine the beam solid angle from the array factor power pattern in the standard way,
\begin{equation}
\Omega_\mathrm{A}=\iint |f(\theta,\phi)|^2\sin\theta\,\mathrm{d}\theta\,\mathrm{d}\phi.
\end{equation}

The tied-array effective area is then
\begin{equation}
A_e =\eta\left(\frac{4\pi\lambda^2}{\Omega_\mathrm{A}}\right),
\end{equation} 
where $ \eta $ is the same frequency and pointing dependent efficiency as in equation~(\ref{eq:tsys}) and $ \lambda $ is the observing wavelength.

Here we note a divergence in the terminology used.
The gain of an aperture array is defined as $ G=4\pi A_e/\lambda^2 = 4\pi\eta/\Omega_\mathrm{A} $ in standard antenna theory. 
We use a different definition (albeit common in radio astronomy), such that the gain is
\begin{equation}
G = \frac{A_e}{2k_\mathrm{B}},
\end{equation}
which relates directly to the system equivalent flux density of the array, $ \mathrm{SEFD}=T_\mathrm{sys}/G $.
In convenient radio astronomy units ($ \mathrm{K\,Jy^{-1}} $, where $ 1\mathrm{\,Jy}=10^{-26} \mathrm{\,W\,m^{-2}\,Hz^{-1}} $) this becomes simply
\begin{equation}
G = \frac{A_e}{2k_\mathrm{B}}\times 10^{-26} \mathrm{\,W\,m^{-2}\,Hz^{-1}},
\end{equation}
where $ k_\mathrm{B} $ is Boltzmann's constant and $ A_e $ is in units of $ \mathrm{m^2} $.

\end{document}